\documentclass[3p,sort&compress]{elsarticle}

\usepackage{lineno,hyperref}
\usepackage[font=normal]{subfig}
\usepackage{amsmath,amssymb,amsfonts,bm,multirow,xcolor,mathtools}
\usepackage[ruled, lined, linesnumbered, commentsnumbered, longend]{algorithm2e}

\newtheorem{thm}{Theorem}

\newdefinition{rmk}{Remark}
\newdefinition{prop}{Proposition}
\newtheorem{definition}{Definition}{}
\newcommand{\proofofref}{}
\newproof{zproofof}{Proof of \proofofref}
\newenvironment{proofof}[1]
{\renewcommand{\proofofref}{#1}\zproofof}
{\endzproofof}

\makeatletter
\newcommand{\Rmnnum}[1]{\expandafter\@slowromancap\romannumeral#1@}

\makeatother

\journal{International Journal of Nonlinear and Robust Control}

\begin{document}

\begin{frontmatter}

\title{Distributed Extended Object Tracking Information Filter Over Sensor Networks\tnoteref{mytitlenote}}


\author[mymainaddress,mysecondaryaddress]{Zhifei~Li\corref{mycorrespondingauthor}}
\cortext[mycorrespondingauthor]{Corresponding author}
\ead{lee@seu.email.cn,lizhifei17@nudt.edu.cn}

\author[mythirdaddress]{Yan~Liang}
\ead{liangyan@nwpu.edu.cn}

\author[mythirdaddress]{Linfeng~Xu}
\ead{xulinf@gmail.com}

\author[mymainaddress]{Shuli~Ma}
\ead{shulima63@163.com}

\address[mymainaddress]{School of Space Information, Space Engineering University, Beijing 101400, China}
\address[mysecondaryaddress]{College of Electronic Engineering, National University of Defense Technology, Hefei 230037, China}
\address[mythirdaddress]{Laboratory of Information Fusion
	Technology, School of Automation, Northwestern Polytechnical University,
	Xi’an 710072, China}

\begin{abstract}
This work aims to design a distributed extended object tracking (EOT) system over a realistic network, where both the extent and kinematics are required to retain consensus within the entire network. To this end, we resort to the multiplicative error model (MEM) that allows the extent parameters of perpendicular axis-symmetric objects to have individual uncertainty. To incorporate the MEM into the information filter (IF) style, we use the moment-matching technique to derive two pair linear models with only additive noise. The separation is merely in a fashion, and the cross-correlation between states is preserved as parameters in each other's model. As a result, the closed-form expressions are transferred into an alternating iteration of two linear IFs. With the two models, a centralized IF is proposed wherein the measurements are converted into a summation of innovation parts. Later, under a sensor network with the communication nodes and sensor nodes, we present two distributed IFs through the consensus on information and consensus on measurement schemes, respectively. Moreover, we prove the estimation errors of the proposed filter are exponentially bounded in the mean square. The benefits are testified by numerical experiments in comparison to state-of-the-art filters in literature.
\end{abstract}


\end{frontmatter}

\section{Introduction}
\label{sec:introduction}
Detecting and tracking moving objects is crucial for many applications such as autonomous driving, region surveillance, and situation awareness \cite{MEOT}. To this end, scan-based sensors such as Radio Detection And Ranging (RADAR) and Light Detection And Ranging (LIDAR) scan the environment and receive the noisy measurement to estimate current and future states. Under such conditions, a sensor network containing multiple nodes will increase the range of surveillance region and tracking performance, especially if the nodes have narrow field-of-view (FoV) and limited sensing distance \cite{WSN21,overview18,overview19}. In general, sensor network involves two types of architectures: centralized and distributed. In centralized system, all data from entire network are gathered into the fusion center to yield an optimal state estimate among all tracking architectures. However, it faces the data congestion issue, especially when the size of network is large. Moreover, the fusion service may be suspended or even denied if the fusion center fails (e.g., suffers from a network attack). In contrast, the distributed system has better scalability and is robust to the node and/or communication link failures \cite{Mallick19}. 
 
\begin{figure}[t]
	\centering
	\subfloat[Example of the point object tracking. A sensor receives single measurement (purple cross) from scattering source (orange dot). Based on the measurement, the objective of a tracker is to determine the kinematic state (e.g., position and velocity) of the object.]{%
		\label{POT}
		\includegraphics[width=0.45\textwidth]{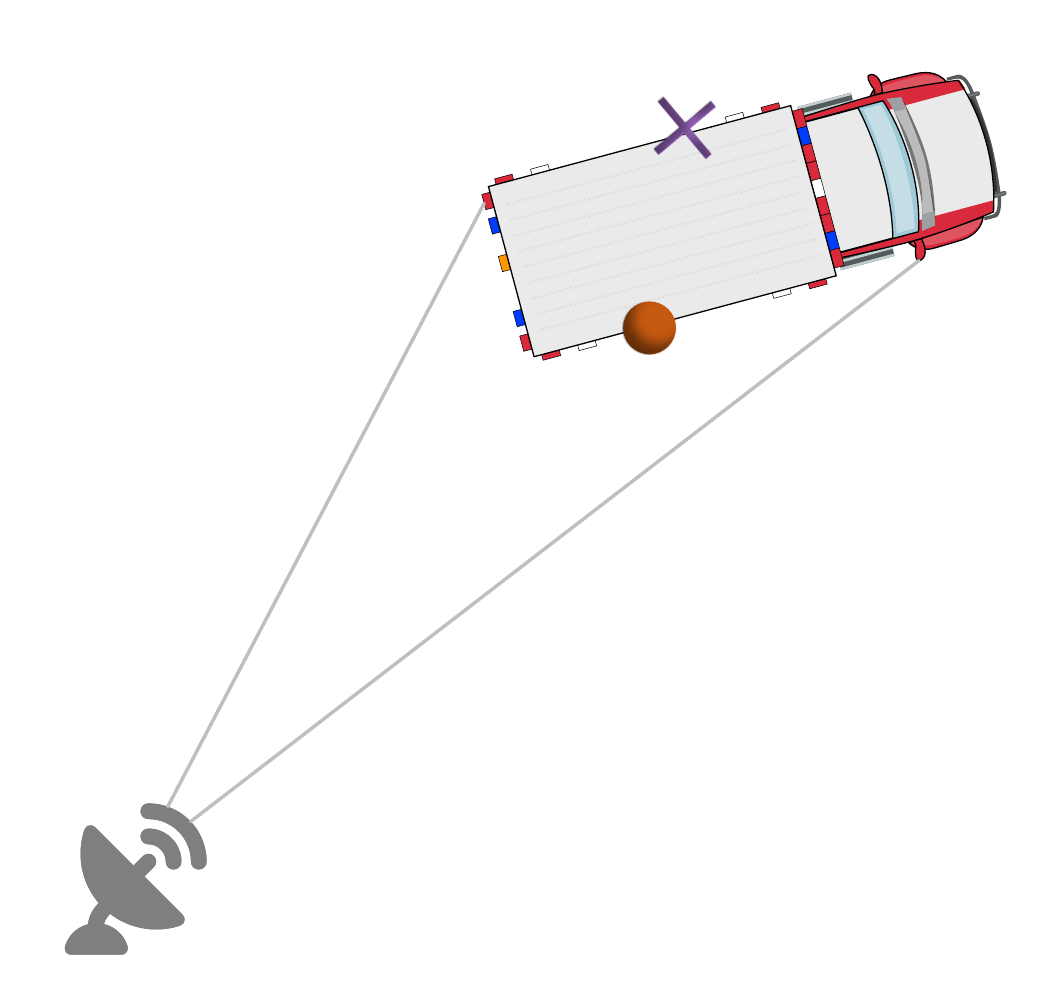}}
	\hfill
	\subfloat[Example of the extended object tracking. A sensor receives multiple measurements (purple cross) from scattering sources (orange dots). Based on these measurements, the objective of a tracker is to jointly determine the kinematic state (e.g., position and velocity) and extent (e.g., size and orientation) of the object.]{\label{EOT}\includegraphics[width=0.45\textwidth]{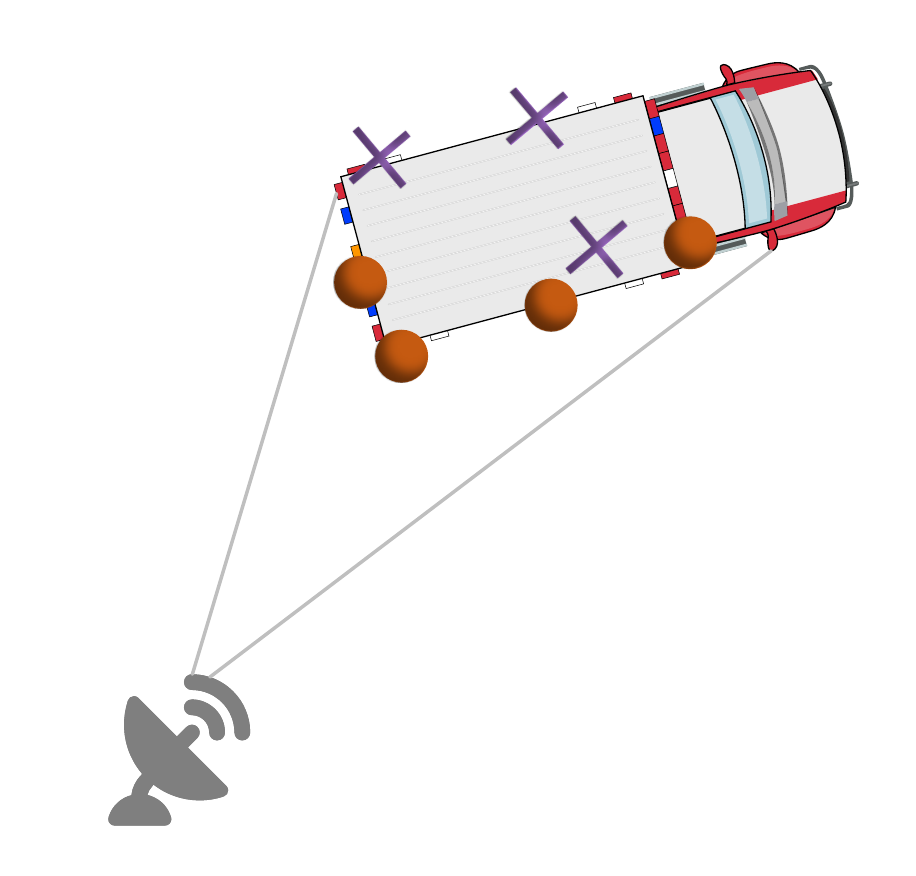}}
	\caption{Two types of object tracking scenarios.}
	\label{fig:tracking}
\end{figure}

Traditional distributed tracking filters consider objects to be tracked as point sources (see Fig. \ref{POT} for an illustration), i.e., their extent is assumed to be neglectable in comparison with sensor resolution. In this context, each node receives at most a single measurement per time step. The most common methods on the point object tracking include average consensus \cite{consensus20,consensus21} and diffusion scheme \cite{diffusion10,diffusion12}. These techniques use the consensus or diffusion steps to provide a consensus estimate within the entire network. To give a solution for the cases where the state-space models contain unknown parameters, some methods are proposed based on the variational Bayesian inference \cite{VB,VB16}. Besides, several approaches with varying focus are available in the setting of asynchronous information fusion \cite{Asy20}, event-triggered mechanism for sensor resource management \cite{event}, and reliable system design under network attacks \cite{attack20,attack21}, etc. 

Due to advances in modern sensors, the original belief that a sensor's resolution is lower than the spatial extent of an object has become obsolete \cite{b1,b2}. In this case, a sensor receives multiple measurements from different scattering sources on the object during a detection process (see Fig. \ref{EOT} for an illustration). This arises a so-called extended object tracking (EOT) problem that estimates kinematic state and extent parameters simultaneously \cite{b3}. The past few years have witnessed various approaches on how to describe the extent, such as the rectangle \cite{b4}, ellipse modeled via the random matrix (RM) \cite{b5,b6}, axis-symmetric shapes modeled via the multiplicative error model (MEM) \cite{b7,b8}, or arbitrary shapes, modeled via the random hyper-surface model (RHM) \cite{b9}, Gaussian process (GP) \cite{b10,b11,b12}, B-splines \cite{b13}, or level-set RHM \cite{b14}. 

To introduce the multi-sensor cooperative tracking manner into the EOT realm, some state-of-the-art approaches have been developed. In \cite{b18,b19}, G. Vivone et al. proposed a centralized EOT filter to simultaneously estimate the kinematics and extent. To overcome the intrinsic demerits in the centralized system, a distributed EOT filter was given by minimizing the weighted Kullback-Leibler divergence in \cite{b20}. Following this, Liu et al. provided an extension for asynchronous sensors \cite{b21}, where the compressed Gaussian mixture approximations of local posterior functions were fused. In \cite{b22}, a distributed variational Bayesian filter was proposed to estimate the unknown states and noise covariance wherein the alternating direction method of multipliers (ADMM) technique was used to give the constraint-based consensus. Besides these RM-based filters \cite{b20,b21,b22}, the work for fusing measurements from multiple sensors based on the MEM model was given in \cite{DEOT}.

\subsection{Motivation}

The limitation on the existing distributed EOT filters \cite{b20,b21,b22,DEOT} is twofold. On the one hand, these filters follow a common prerequisite that an object is within all nodes' FoV regardless of scan time. However, in a real-world sensor network, the deployed node has limited scan angle and sensing distance due to constraints on resources and hardware devices. In other words, an object is usually seen by only a few of nodes at each scan time, even though the full set of nodes view the object at any time. In such a scenario, when the nodes fail to get the measurements, their estimates are always far away from the ground truth. Even if the consensus operation eliminates the difference between nodes, the estimates on the nodes without measurements still have an opposite effect on the final consensus result. On the other hand, the RM-based filters determine the temporal evolution of extent via a single scalar value, while the degree of uncertainty on extent parameters may be different over time. Moreover, the RM model involves a special structure (i.e., the process noise covariance is related to the object's extent) to ensure a recursive operation, which limits other kinematics, such as yaw rate, being easily merged into the state vector. With the above analysis, a computationally efficient and actually available distributed EOT system is still lacking. 

\subsection{Results and organization}

In this work, we endeavor to design a distributed EOT filter over a realistic sensor network. The extent is described by the MEM model proposed in \cite{b8} that partitions the extent into a 3-D vector including semi-axes and orientation. Based on the MEM model, we derive two distributed information filters to yield a consensus estimate on both the extent and kinematics. The main contributions are as follows.
\begin{enumerate}
	\item We derive two pair linear measurement models with only additive noise through the first-order Taylor series expansion followed by the moment-matching technique. The two models preserve the cross-correlation between states of the original MEM model. More importantly, they pave the way for the related filter to suit the standard IF framework. 
	\item This work presents a compact centralized information filter based on the two models, where the massive measurements are converted into a summation form of innovation parts. The centralized filter is suitable in a small-size network, and here it also serves as a benchmark to examine the performance of the corresponding distributed filter.
	\item We give two types of distributed information filters by using \textit{Consensus on Information} (CI) and \textit{Consensus on Measurement} (CM) schemes, respectively. The two filters yield the consensus estimates on both the extent and kinematics. Moreover, we investigate that the estimation errors of the proposed filter are exponentially bounded in the mean square. 
\end{enumerate}

The rest of the work is organized as follows. Section \ref{sec:problem} gives a brief problem formulation. Section \ref{sec:sepe} presents two separate measurement models. Section \ref{sec:centralized} gives a centralized filter and Section \ref{sec:distributed}  presents two corresponding distributed filters. Section \ref{sec:analysis} explores the stability of the proposed distributed filter. The proposed filters are demonstrated by numerical examples in Section \ref{sec:simulation}. Section \ref{sec:conclusion} concludes this work.

\textit{Notation}: For clarity, we use italics to denote scalar quantities and boldface for vectors and matrices. Let the notation $X X^{\mathsf{T}} = X\left(\star\right)^{\mathsf{T}}$ and $X^{\mathsf{T}}YX = \left(\star\right)^{\mathsf{T}}YX$. We use ``$:=$'' to define a quantity and $(\cdot)^{\mathsf{T}}$ denotes the transpose of a matrix/vector. The $n$-th dimensional identity matrix is denoted by $\mathbf I_n$. Operation $\mathrm{col}(\mathbf A_i)_{i \in \mathcal N}$, where $\mathcal N$ is a finite set, denotes stacking $\mathbf A_i$ on top of each other to form a column matrix.

\section{Problem Formulation}
\label{sec:problem}
Here, we consider the MEM model given in \cite{b8} to achieve the centralized and distributed information filters over a sensor network \cite{b25}. The network consisting of two types of nodes is deployed over a geographic region. Therein, the \textit{Communication} nodes process local measurements as well as communicate with neighboring nodes, while \textit{Sensor} nodes also have sensing capabilities. The network is served as a case of a real architecture, where the communication nodes represent those nodes that cannot detect objects due to the limited FoV. The network is denoted by triplet $(\mathcal{S, C, A})$, where $\mathcal{S}$ is the set of sensor nodes, $\mathcal{C}$ is the set of communication nodes, $\mathcal{G = S\cup C}$, $\mathcal{A \subseteq G \times G}$ is the set of edges such that $(s, j) \in \mathcal{A}$ if node $s$ can communicate with $j$. For each node $s \in \mathcal{G}$, $\mathcal G^s$ denotes the set of its neighboring nodes (including $s$ itself), i.e., $\mathcal G^s := \{j: (j,s) \in \mathcal{A}\}$, and let $\mathcal G^s \setminus \{s\}$ be the set of its neighboring nodes (excluding $s$ itself). Let $\mathcal Y_{k,s} = \{\bm y_{k,s}^{i} \}^{n_{k,s}}_{i=1}$ denotes a set of $n_{k,s}$ independent 2-D position measurements on node $s \in \mathcal{S}$ at time $k$, and the collective data from all sensor nodes is denoted as $\mathcal Y_k = \{\mathcal Y_{k,s}\}_{s \in \mathcal{S}}$.

\begin{figure}[htbp]
	\centering\includegraphics[width=0.45\textwidth]{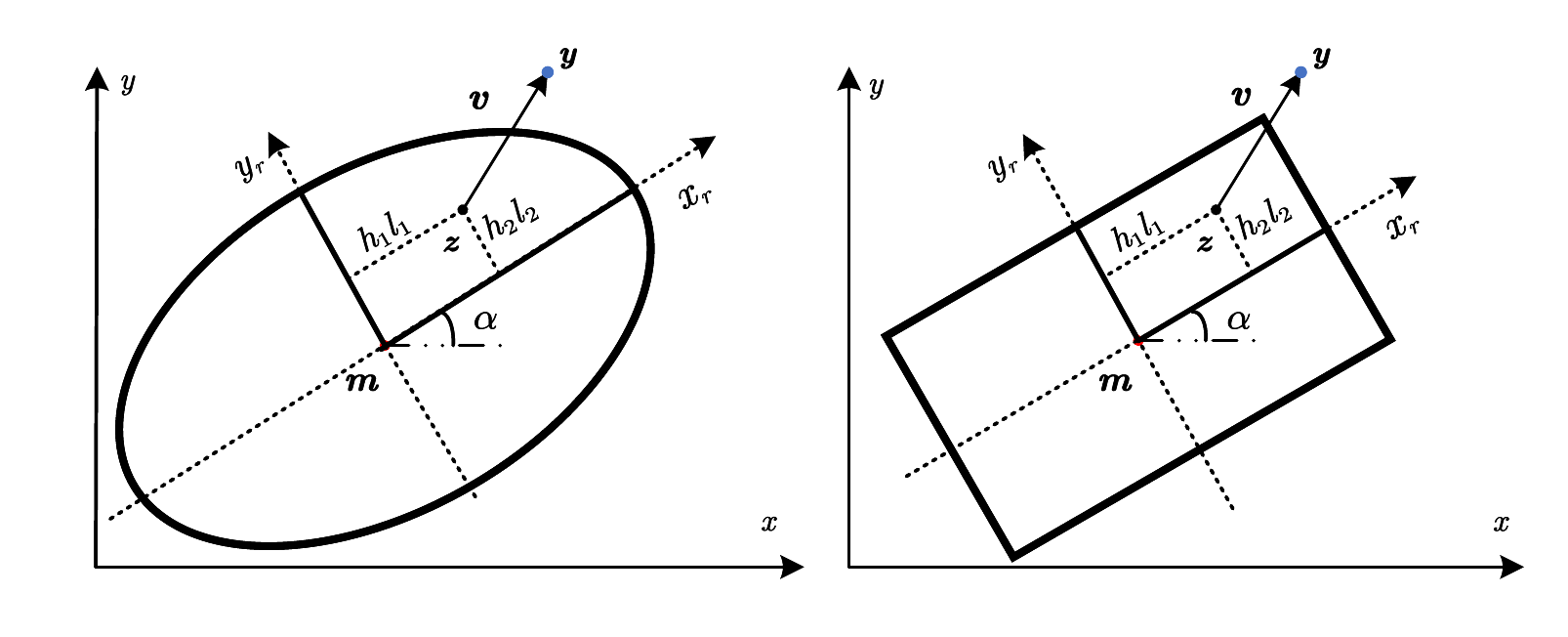}
	\caption{{Illustration of the measurement model. Here, the time index $k$, measurement index ${i}$, and sensor node index $s$ are omitted for brevity if no ambiguity arises. The centroid position of the object is $\bm m = [m_1, m_2]^{\mathsf{T}}$, and its extent is denoted as $\bm p = [\alpha, l_1,l_2]^{\mathsf{T}}$. By counterclockwise rotating an angle $\alpha$ (i.e., the orientation) along the $x$-axis, we get the reference coordinates $x_r$-$y_r$. The scattering source $\bm z$ is determined by parameters $\bm p$, $\bm m$, and multiplicative noise $\bm h = [h_1,h_2]^{\mathsf{T}}$. The measurement $\bm y$ is generated via the source $\bm z$ plus a Gaussian measurement noise $\bm v$. The illustration gives an intuitive view that the measurement model is feasible to describe the perpendicular axis-symmetric shapes, such as an ellipse or a rectangular.}} 
	\label{fig1}
\end{figure}

Next, we first review briefly the MEM model.
\begin{enumerate}[1.]
	\item State Parameterization
\end{enumerate}
The kinematic state $\bm x_k$ of an object at time $k$ 
\begin{equation} \label{eq1}
	\bm x_k = [\bm m_k^{\mathsf{T}}, \bm {\dot{m}}_k^{\mathsf{T}}, \cdots]^{\mathsf{T}}
\end{equation}
includes the position $\bm m_k$, velocity $\bm {\dot{m}}_k$, and interested quantities such as acceleration. The extent at time $k$ 
\begin{equation} \label{eq2}
	\bm p_k = [\alpha_k, l_{k,1}, l_{k,2}]^{\mathsf{T}} \: \in \mathbb{R}^3
\end{equation}
contains the orientation $\alpha_k$, which denotes the counterclockwise rotation angle along the $x$-axis, and the semi-axes lengths $l_{k,1}$ and $l_{k,2}$.
\begin{enumerate}[2.]
	\item Measurement Model
\end{enumerate}
At time $k$, the ${i}$-th measurement $\bm y_{k,s}^{i}$ on sensor node $s\in\mathcal{S}$ is modeled as 
\begin{equation} \label{eq3}
	\bm y_{k,s}^{i}\! = \!\mathbf{H} \bm x_k + \underbrace{\begin{bmatrix}
			\cos{\alpha_k} & -\sin{\alpha_k}  \\ \sin{\alpha_k} & \cos{\alpha_k} 
		\end{bmatrix} \; \begin{bmatrix}
			l_{1,k} & 0 \\ 0 & l_{2,k}
	\end{bmatrix}}_{:= \mathbf S_k} \; \underbrace{\begin{bmatrix}
			h_{k,1}^{i} \\ h_{k,2}^{i}
	\end{bmatrix}}_{:= \bm h_{k,s}^{i}} + \bm v_{k,s}^{i}
\end{equation}
where $\mathbf H = \left[ \mathbf I_2 \; \bm 0 \right]$ is the measurement matrix, $\mathbf S_k$ compacts the orientation and size for the considered object, multiplicative noise $\bm h_{k,s}^{i}$ with covariance $\mathbf C^{h}$ guarantees that any scattering source $\bm z_{k,s}^{i}$ lies on the boundary or interior of an object (see Fig. \ref{fig1}), and $\bm v_{k,s}^{i}, \bm v_{k,j}^{i}, \cdots$ are uncorrelated zero-mean Gaussian noises with covariances $\mathbf C_s^v \delta_{sj}$ if $s=j$, $\delta_{sj}=1$, and $\delta_{sj}=0$, otherwise.     
\begin{enumerate}[3.]
	\item Evolution Models
\end{enumerate}
The following formulas show the temporal evolution models for the kinematics and extent, respectively
\begin{equation} \label{eq4}
	\bm x_{k+1} = \mathbf F_k^x \bm x_k + \bm w_k^x
\end{equation} 
\begin{equation} \label{eq5}
	\bm p_{k+1} = \mathbf F_k^p \bm p_k + \bm w_k^p
\end{equation}
where $\mathbf F_k^x$ and $\mathbf F_k^p$ are state transition matrices, and $\bm w_k^x$ and $\bm w_k^p$ are zero-mean Gaussian process noises with covariances $\mathbf C_w^x$ and $\mathbf C_w^p$, respectively. The two transition matrices need to be determined according to the actual motion mode and body structure, e.g., for a rigid object with nearly constant velocity, 
\begin{equation*}
	\mathbf F_k^x = \begin{bmatrix}
		1  & 0 & \mathrm{T} & 0 \\ 0 & 1 & 0 & \mathrm{T} \\ 0 & 0 & 1 & 0 \\ 0 & 0 & 0 & 1 
	\end{bmatrix}, \qquad \mathbf F_k^p = \mathbf I_3
\end{equation*}
where $\mathrm{T}$ is the scan time.

This work takes the MEM model in \cite{b22} as a basis due to the following merits: (1) it reduces the extent of perpendicular axis-symmetric objects to a 3-D vector $\bm p_k$, (2) the dynamic models \eqref{eq4} and \eqref{eq5} are treated independently, allowing for other states to be incorporated into the state vector, and (3) the uncertainty of the extent is determined by the semi-axes and orientation, not a single scalar value.
\section{Separation of the coupled kinematics and extent}
\label{sec:sepe}
Here, we utilize the strength of IF in multi-sensor control to design the centralized/distributed filter. For this purpose, a linear state-space model with only additive noise is necessary to suit an IF style. However, the model \eqref{eq3} including multiplicative noise is highly nonlinear. Hence, the top priority is to construct two separate linear measurement models with only additive noise w.r.t $\bm x_k$ and $\bm p_k$. These models are then fed to the IF framework to determine the unknown states. 

The measurements $\{\bm y_{k,s}^{i} \}^{n_{k,s}}_{i=1}$ at each sensor node $s \in \mathcal S$ are processed in a sequential way. Let $ \hat{\bm x}_{k}^{[i-1]}$, $\hat{\bm p}_{k}^{[i-1]}$, $\mathbf C_{k}^{x[i-1]}$ and $\mathbf C_{k}^{p[i-1]}$ denote the estimates for the kinematics $\bm x_{k}$ and extent $\bm p_{k}$ together with their covariances at the ${[i-1]}$-th sequential operation. The node $s$  processes $\bm y_{k,s}^{i}$ to obtain the updated estimates $ \hat{\bm x}_{k}^{[i]}$, $\hat{\bm p}_{k}^{[i]}$, $\mathbf C_{k}^{x[i]}$ and $\mathbf C_{k}^{p[i]}$. Next, we focus on constructing two pair linear measurement models w.r.t $\bm x_k$ and $\bm p_k$, respectively.
\begin{prop} \label{kineup}
	The measurement model related to $\bm x_k$ is 
	\begin{equation} \label{eq6}
		\bm y_{k,s}^{i} \approx \mathbf{H} \bm x_k + \bm v_{k,s}^{x[i]}
	\end{equation}
	where $\bm v_{k,s}^{x[i]}$ is the equivalent noise about $\bm x_k$ with $\mathbb{E} (\bm v_{k,s}^{x[i]}) = \bm 0 $ and $\mathrm{Cov} (\bm v_{k,s}^{x[i]}) = \mathbf R_{k,s}^{x[i]} := \mathbf C^{\mathrm{\Rmnnum{1}}} + \mathbf C^{\mathrm{\Rmnnum{2}}} + \mathbf C_s^v$. The terms $\mathbf C^{\mathrm{\Rmnnum{1}}}$ and $\mathbf C^{\mathrm{\Rmnnum{2}}}$ are
	\begin{equation} \label{eq7}
		\mathbf C^{\mathrm{\Rmnnum{1}}} = \hat{\mathbf S}_k^{[i-1]} \mathbf C^h \left(\hat{\mathbf S}_k^{[i-1]}\right)^{\mathsf{T}}
	\end{equation}
	
	\begin{equation} \label{eq8}
		\underbrace{[\epsilon_{mn}]}_{\mathbf C^{\mathrm{\Rmnnum{2}}}} = \mathrm{tr} \left\{ \mathbf C_k^{p[i-1]} \left(\hat{\mathbf J}_{n,k}^{[i-1]}\right)^{\mathsf{T}}\mathbf C^h \hat{\mathbf J}_{m,k}^{[i-1]}\right\}
	\end{equation}
	for $m,n \in \{1,2\}$. The quantities $\hat{\mathbf J}_{1,k}^{[i-1]}$  and $\hat{\mathbf J}_{2,k}^{[i-1]}$ are the Jacobian matrices of the first row $\mathbf S_{1,k}$ and second row $\mathbf S_{2,k}$ of $\mathbf S_k$ around the ${[i-1]}$-th extent estimate $\hat{\bm p}_k^{[i-1]}$, respectively.
\end{prop}

\begin{proofof}[Proof of Proposition \ref{kineup}.]
	Since the true extent $\bm p_k$ is unknown in the shape matrix $\mathbf S_k$, we take a first-order Taylor series	approximation on $\mathbf S_k \bm h_{k,s}^{i}$ in \eqref{eq3} around the ${[i-1]}$-th extent estimate $\hat{\bm p}_k^{[i-1]}$ and retain $\bm h_{k,s}^{i}$ as a random variable to yield
	\begin{equation} \label{eq9}
		\mathbf S_k \bm h_{k,s}^{i} \approx \underbrace{\hat{\mathbf{S}}_k^{[i-1]} \boldsymbol{h}_{k,s}^{i}}_{\mathrm{I}}+\underbrace{\left[\begin{array}{c}
				\left(\boldsymbol{h}_{k,s}^{i}\right)^{\mathsf{T}} \hat{\mathbf{J}}_{1,k}^{[i-1]} \\
				\left(\boldsymbol{h}_{k,s}^{i}\right)^{\mathsf{T}} \hat{\mathbf{J}}_{2,k}^{[i-1]}
			\end{array}\right]\left(\bm{p}_k-\hat{\bm{p}}_k^{[i-1]}\right)}_{\mathrm{II}}.
	\end{equation}
	Substituting \eqref{eq9} into \eqref{eq3}, the residual covariance about $\bm y_{k,s}^{i}$ in \eqref{eq3} is calculated as  
	\begin{equation} \label{eq10}
		\mathbf C_{k,s}^{y[i]} = \mathbf H \mathbf C_k^{x[i-1]} \mathbf H^{\mathsf{T}} + \mathbf C^{\mathrm{\Rmnnum{1}}} + \mathbf C^{\mathrm{\Rmnnum{2}}} + \mathbf C_s^v,
	\end{equation} 
	and \eqref{eq3} is rewritten as \eqref{eq6}.
	The proof is complete. 
\end{proofof} 
Note that the quantities $\mathbf C^{\mathrm{\Rmnnum{1}}}$ and $\mathbf C^{\mathrm{\Rmnnum{2}}}$ in \eqref{eq10} are treated as constant terms at the $[i]$-th sequential operation since they are calculated based on the former estimate $\hat{\bm p}_k^{[i-1]}$.

As pointed out in \cite{b8,b27}, a pseudo-measurement using 2-fold Kronecker product is required to update the extent. The ${i}$-th pseudo-measurement $\mathbf Y_k^{i}$ is given as  
\begin{equation} \label{eq11}
	\mathbf Y_{k,s}^{i} = \mathbf F \left( (\bm y_{k,s}^{i} - \mathbf H \hat{\bm x}_k^{[i-1]})\otimes (\bm y_{k,s}^{i} - \mathbf H \hat{\bm x}_k^{[i-1]}) \right)
\end{equation} 
with 
\begin{equation} \label{eq12}
	\mathbf F = \begin{bmatrix}
		1  & 0 & 0 & 0 \\ 0 & 0 & 0 & 1 \\ 0 & 1 & 0 & 0
	\end{bmatrix}.
\end{equation}

Based on \eqref{eq11}, the following proposition \ref{exteup} gives the measurement model w.r.t the extent $\bm p_k$.
\begin{prop} \label{exteup}
	The measurement model related to $\bm p_k$ is
	\begin{equation} \label{eq13}
		\mathbf Y_{k,s}^{i} \approx \hat{\mathbf M}_k^{[i-1]} \bm p_k + \bm v_{k,s}^{p[i]}
	\end{equation}
	where $\bm v_{k,s}^{p[i]}$ is the equivalent noise about $\bm p_k$ with 
	\begin{equation*}
		\mathbb E (\bm v_{k,s}^{p[i]})=\bar{\bm v}_{k,s}^{p[i]}:=\mathbf F \mathrm{vect} \left(\mathbf C_{k,s}^{y[i]}\right) - \hat{\mathbf M}_k^{[i-1]} \hat{\bm p}_k^{[i-1]},
	\end{equation*}
	\begin{equation*}
		\begin{split}
			\mathrm{Cov} (\bm v_{k,s}^{p[i]}) = \mathbf R_{k,s}^{p[i]} := & \mathbf F \left( \mathbf C_{k,s}^{y[i]}\otimes \mathbf C_{k,s}^{y[i]} \right) (\mathbf F + \tilde{\mathbf F})^{\mathsf{T}}\\& - \hat{\mathbf M}_k^{[i-1]} \mathbf C_k^{p[i-1]} \left(\hat{\mathbf M}_k^{[i-1]}\right)^{\mathsf{T}},
		\end{split}
	\end{equation*}
	with 
	\begin{equation} \label{eq14}
		\tilde{\mathbf F} = \begin{bmatrix}
			1 & 0 & 0 & 0 \\ 0 & 0 & 0 & 1 \\ 0 & 0 & 1 & 0
		\end{bmatrix},
	\end{equation}
	and 
	\begin{equation} \label{eq15}
		\hat{\mathbf M}_k^{[i-1]} = \begin{bmatrix}
			2 \hat{\mathbf S}_{1,k}^{[i-1]} \mathbf C^h \hat{\mathbf J}_{1,k}^{[i-1]} \\
			2 \hat{\mathbf S}_{2,k}^{[i-1]} \mathbf C^h \hat{\mathbf J}_{2,k}^{[i-1]} \\
			\hat{\mathbf S}_{1,k}^{[i-1]} \mathbf C^h \hat{\mathbf J}_{2,k}^{[i-1]} +
			\hat{\mathbf S}_{2,k}^{[i-1]} \mathbf C^h \hat{\mathbf J}_{1,k}^{[i-1]}
		\end{bmatrix}.
	\end{equation}
\end{prop}

\begin{proofof}[Proof of Proposition \ref{exteup}.]
	See Appendix. 
\end{proofof}
Note that $\hat{\mathbf M}_k^{[i-1]}$ corresponds to the measurement matrix w.r.t $\bm p_k$, and it is treated as constant terms at the $[i]$-th sequential operation. 
\begin{rmk}
	\begin{itemize}
		\item The dual linear state-space model is built using the models {\eqref{eq4} , \eqref{eq6}} and {\eqref{eq5}, \eqref{eq13}} such that each node retains a traditional Kalman or IF style.
		\item The models \eqref{eq6} and \eqref{eq13} are separated in a fashion, but the cross-correlation about $\bm x_k$ and $\bm p_k$ is still remained in each other's model. In this way, the joint estimation is merged into an iterative implementation of two linear filters. 
		\item The models \eqref{eq6} and \eqref{eq13} only utilize the first-order Taylor series expansion in \eqref{eq9}, omitting the higher-order terms. On the one hand, they avoid some complex calculations, such as the Hessian matrix. On the other hand, \eqref{eq13} is only reliant on the first-order expansion as a prerequisite, otherwise \eqref{eq13} cannot be derived.   
	\end{itemize}    
\end{rmk}
\section{Centralized Extended Object Tracking Filter}
\label{sec:centralized}
 Here, we resort to the two pair models {\eqref{eq4}, \eqref{eq6}} and {\eqref{eq5},  \eqref{eq13}} to derive a centralized EOT (CEOT) information filter. Information filter updates, instead of the estimate $\hat{\bm x}_{k}^{[i]} (\hat{\bm p}_{k}^{[i]})$ and its error covariance $\mathbf C_{k}^{x[i]} (\mathbf C_{k}^{p[i]})$, the information matrix $\mathbf{\Omega}^{x[i]}_{k} := \left(\mathbf C_{k}^{x[i]}\right)^{-1}$ and information vector $\hat{\bm q}^{x[i]}_{k} := \left(\mathbf C_{k}^{x[i]}\right)^{-1}\hat{\bm x}_{k}^{[i]}$ \cite{b26}. 

In CEOT filter, the fusion center sequentially processes all measurements $\mathcal Y_k = \{\mathcal Y_{k,s}\}_{s \in \mathcal{S}}$ from all sensor nodes to obtain an estimate. Assume that there are $n_k$ measurements, for clarity, on each sensor node $s \in \mathcal S$ at time $k$. Given the $[i-1]$-th estimates $ \hat{\bm q}^{x[i-1]}_{k}$, $\hat{\bm q}^{p[i-1]}_{k}$, $\mathbf{\Omega}^{x[i-1]}_{k}$ and $\mathbf{\Omega}^{p[i-1]}_{k}$, the center processes the measurement set $\{\bm y_{k,s}^{i}\}_{s \in \mathcal{S}}$ to give the $[i]$-th estimates. Notice that the notation $(\cdot)_{k}^{[0]}$ is the corresponding predicted estimates at time $k$.

Define the central measurement, measurement matrix, and noise covariance with regard to the measurement set $\{\bm y_{k,s}^{i}\}_{s \in \mathcal{S}}$ as 
\begin{equation}\label{eq16}
	\left \{ \begin{lgathered} 
		\bm y_{k,\mathrm{c}}^{i} := \mathrm {col} (\bm y_{k,1}^{i},\bm y_{k,2}^{i},\cdots,\bm y_{k,\lvert \mathcal S \rvert}^{i}) \\
		\mathbf H_{\mathrm{c}}^{[i]} := [\mathbf H;\mathbf H;\cdots;\mathbf H]\\
		\mathbf R_{k,\mathrm{c}}^{x[i]} := \mathrm {diag} (\mathbf R_{k,1}^{x[i]},\mathbf R_{k,2}^{x[i]},\cdots,\mathbf R_{k,\lvert \mathcal S \rvert}^{x[i]})
	\end{lgathered} \right. 
\end{equation}
where the subscript ``c'' denotes ``central'', and $\lvert \mathcal S \rvert$ is the cardinality of $\mathcal S$.

Define the central measurement, measurement matrix, and noise covariance with regard to the pseudo-measurement set $\{\tilde{\mathbf Y}_{k,s}^{i}\}_{s \in \mathcal{S}}$ as 
\begin{equation}\label{eq17}
	\left \{ \begin{lgathered} 
		\tilde{\mathbf Y}_{k,s}^{[i]} := \mathbf Y_{k,s}^{i}-\mathbf F \mathrm{vect} \left(\mathbf C_{k,s}^{y[i]}\right) + \hat{\mathbf M}_{k}^{[i-1]} \hat{\bm p}_{k}^{[i-1]} \\
		\tilde{\mathbf Y}_{k,\mathrm{c}}^{[i]} := \mathrm {col} (\tilde{\mathbf Y}_{k,1}^{[i]},\tilde{\mathbf Y}_{k,2}^{[i]},\cdots,\tilde{\mathbf Y}_{k,\lvert \mathcal S \rvert}^{[i]})\\
		\mathbf M_{k,\mathrm{c}}^{[i]} := [\hat{\mathbf M}_{k}^{[i-1]};\hat{\mathbf M}_{k}^{[i-1]};\cdots;\hat{\mathbf M}_{k}^{[i-1]}]\\
		\mathbf R_{k,\mathrm{c}}^{p[i]} = \mathrm {diag} (\mathbf R_{k,1}^{p[i]},\mathbf R_{k,2}^{p[i]},\cdots,\mathbf R_{k,\lvert \mathcal S \rvert}^{p[i]})
	\end{lgathered} \right.. 
\end{equation}

Introducing the noise information matrices $\mathbf W_w^x := (\mathbf C_w^x)^{-1}$, $\mathbf W_w^p := (\mathbf C_w^p)^{-1}$, $\mathbf V_{k,s}^{x[i]} := (\mathbf R_{k,s}^{x[i]})^{-1}$ and $\mathbf V_{k,s}^{p[i]} := (\mathbf R_{k,s}^{p[i]})^{-1}$, CEOT filter based on \eqref{eq16} and \eqref{eq17} consists of two steps:\\
(1) Measurement Update Step (Correction)
\begin{subequations} \label{eq18}
	\begin{equation}\label{eq18a}
		\hat{\bm q}_k^{x[i]} = \hat{\bm q}_k^{x[i-1]} + \sum_{s \in \mathcal{S}} \mathbf H^{\mathsf{T}} \mathbf V_{k,s}^{x[i]} \bm{y}_{k,s}^{i}
	\end{equation}
	\begin{equation}\label{eq18b}
		\mathbf{\Omega}_k^{x[i]} = \mathbf{\Omega}_k^{x[i-1]}+\sum_{s\in\mathcal S}\mathbf H^{\mathsf{T}} \mathbf V_{k,s}^{x[i]} \mathbf{H},
	\end{equation}
\end{subequations}

\begin{subequations} \label{eq19}
	\begin{equation} \label{eq19a}
		\hat{\bm q}_k^{p[i]}=\hat{\bm q}_k^{p[i-1]}+\sum_{s \in \mathcal{S}} \left(\hat{\mathbf M}_{k}^{[i-1]}\right)^{\mathsf{T}} \mathbf V_{k,s}^{p[i]} \tilde{\mathbf Y}_{k,s}^{[i]}
	\end{equation}
	\begin{equation}  \label{eq19b}
		\mathbf{\Omega}_k^{p[i]} = \mathbf{\Omega}_k^{p[i-1]}+\sum_{s \in \mathcal{S}}\left(\hat{\mathbf M}_{k}^{[i-1]}\right)^{\mathsf{T}} \mathbf V_{k,s}^{p[i]} \hat{\mathbf M}_{k}^{[i-1]}.
	\end{equation}
\end{subequations}

It is worth noting that \eqref{eq18} and \eqref{eq19} use the models \eqref{eq6} and \eqref{eq13} to construct a standard IF form wherein the measurements are converted into a summation term of innovation parts (we refer $\mathbf H^{\mathsf{T}} \mathbf V_{k,s}^{x[i]} \mathbf{H}$ and $\mathbf H^{\mathsf{T}} \mathbf V_{k,s}^{x[i]} \bm{y}_{k,s}^{i}$ as innovation part related to $\bm x_k$). This operation further reduces the computational cost in contrast to the original model \eqref{eq3}. \\
(2) Time Update Step (Prediction)\\
After the $n_k$ batch of measurements is processed sequentially, the prediction step is conducted using the IF formulas with
\begin{subequations} \label{eq20}
	\begin{equation} \label{eq20a}
		\hat{\bm q}_{k+1}^{x[0]} = \mathbf{\Omega}_{k+1}^{x[0]}\,\mathbf{F}_k^x\, {\mathbf{\Omega}_k^{x[n_k]}}^{-1}\hat{\bm q}_k^{x[n_k]}
	\end{equation}
	\begin{equation} \label{eq20b}
		\mathbf{\Omega}_{k+1}^{x[0]} = \mathbf{W}_w^x-\mathbf{W}_w^x \mathbf{F}_k^x\left(\mathbf{\Omega}_k^{x[n_k]}+\mathbf{F}_k^{x\mathsf{T}} \mathbf{W}_w^x \mathbf{F}_k^x\right)^{-1} \mathbf{F}_k^{x\mathsf{T}} \mathbf{W}_w^x
	\end{equation}
\end{subequations}

\begin{subequations} \label{eq21}
	\begin{equation} \label{eq21a}
		\hat{\bm q}_{k+1}^{p[0]} = \mathbf{\Omega}_{k+1}^{p[0]}\,\mathbf{F}_k^p\, \left( \mathbf{\Omega}_k^{p[n_k]}\right)^{-1}\hat{\bm q}_k^{p[n_k]}
	\end{equation}
	\begin{equation} \label{eq21b}
		\mathbf{\Omega}_{k+1}^{p[0]} = \mathbf{W}_w^p-\mathbf{W}_w^p \mathbf{F}_k^p\left(\mathbf{\Omega}_k^{p[n_k]}+\mathbf{F}_k^{p\mathsf{T}} \mathbf{W}_w^p \mathbf{F}_k^p\right)^{-1} \mathbf{F}_k^{p\mathsf{T}} \mathbf{W}_w^p.
	\end{equation}
\end{subequations}
The detailed CEOT filter is collected in Algorithm \ref{algorithm1}.  
\begin{algorithm}[h]
	\caption{Centralized EOT (CEOT) Filter} \label{algorithm1}	
	\SetAlgoLined
	\textbf{Initialization:} $\hat{\bm x}_{1}^{[0]}$, $\hat{\bm p}_{1}^{[0]}$, $\mathbf{\Omega}_{1}^{x[0]}$, and $\mathbf{\Omega}_{1}^{p[0]}$ \;
	\For{$k \gets 1,2,\cdots $ \tcp*[h]{scan time} }{
		\textbf{Data:} $\{\bm y_{k,s}^{i} \}^{n_{k}}_{i=1} (s \in \mathcal N)$ \;
		Initialization: $\hat{\bm x}_{k}^{[0]}$, $\hat{\bm p}_{k}^{[0]}$, $\mathbf{\Omega}_{k}^{x[0]}$, and $\mathbf{\Omega}_{k}^{p[0]}$ \;
		
		\For{$i=1,2,\cdots,n_k$ \tcp*[h]{sequential}}{compute $\hat{\bm x}_{k}^{[i]}, \mathbf{\Omega}_{k}^{x[i]}, \hat{\bm p}_{k}^{[i]}, \mathbf{\Omega}_{k}^{p[i]}$ via \eqref{eq18} and \eqref{eq19}}
		\textbf{Output1:} $\hat{\bm{x}}_k \gets \left( \mathbf{\Omega}_k^{x[n_k]}\right)^{-1}\hat{\bm q}_k^{x[n_k]}, \mathbf C_{k}^{x} \gets \left( \mathbf{\Omega}_{k}^{x[n_k]}\right)^{-1}$ \;
		\textbf{Output2:} $\hat{\bm{p}}_k \gets \left( \mathbf{\Omega}_k^{p[n_k]}\right)^{-1}\hat{\bm q}_k^{p[n_k]},\mathbf C_{k}^{p} \gets \left( \mathbf{\Omega}_{k}^{p[n_k]}\right)^{-1}$ \;
		\textbf{Prediction:} compute $\hat{\bm q}_{k+1}^{x[0]},\mathbf{\Omega}_{k+1}^{x[0]},\hat{\bm q}_{k+1}^{p[0]},\mathbf{\Omega}_{k+1}^{p[0]}$ via \eqref{eq20} and \eqref{eq21}	
	}
\end{algorithm}

\begin{figure}[t]
	\centering
	\includegraphics[scale=.6]{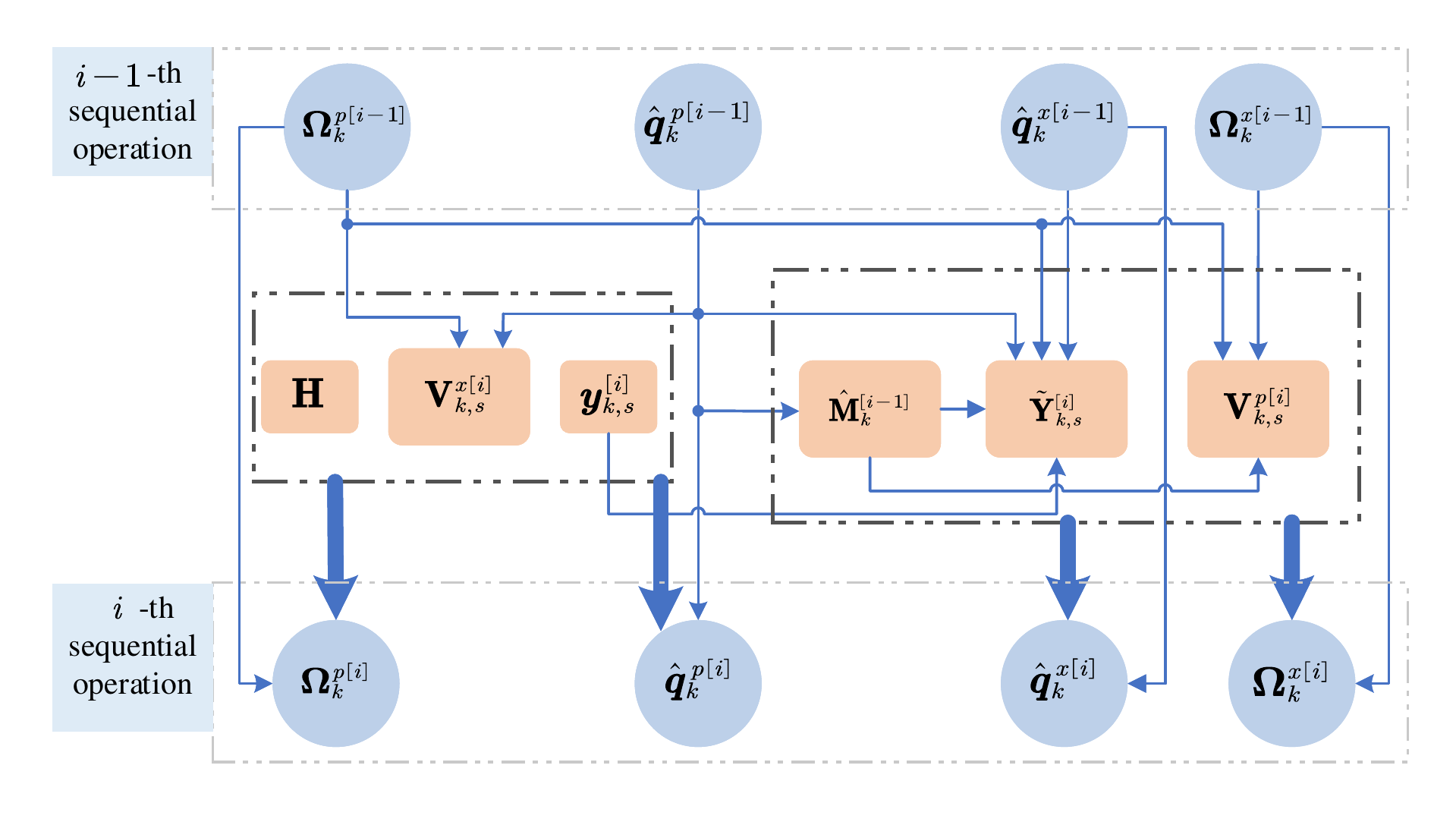}
	\caption{{An illustration of the cross-correlation between $\bm x_k$ and $\bm p_k$.}}
	\label{sequen}
\end{figure}

Notice that using \eqref{eq6} and \eqref{eq13} to achieve the corresponding filters causes that the cross-correlation between $\bm x_k$ and $\bm p_k$ retains in the $[i]$-th and $[i-1]$-th sequential operation. Here, we just take CEOT filter as an example to show the cross-correlation (see Fig. \ref{sequen}).   

\section{Distributed Extended Object Tracking Filters}
\label{sec:distributed}
The distributed tracking system has some appealing advantages over the centralized system for overcoming the communication bandwidth constraints, single-node failure, and congestion of massive data. This section extends CEOT into a distributed scenario. To this end, we apply two schemes, namely consensus on information and consensus on measurement, to achieve consensus estimates among all nodes, respectively. Considering the existence of multiple measurements on sensor nodes, before yielding the final results, the sequential processing technique is required at each time step.
\subsection{Distributed Consensus on Information Filter}
\label{subCI}
Since communication nodes $\mathcal C$ do not have measurements, their local estimates are often erroneous. Error covariance is an indicator to point out the error range of estimates. Hence, a reasonable solution is that the estimate in a local node is appropriately weighted by its information matrix. Then a node with less information about the estimated state will have less weight in the consensus process \cite{b28,b29}. Inspired by the idea, we propose a distributed CI information filter, where each local node receives and transmits the information vector and information matrix to its neighboring nodes (i.e., track-to-track fusion). 

Next, we first define a convex combination (CC) fusion rule. Then, we will show how to incorporate the rule into CI filter.
\begin{definition} [Convex Combination fusion] \label{def1}
	Assume that the network $\mathcal G$ is strongly connected and undirected, given the consensus weights $\{\pi ^{s,j}\}$ (e.g., the Metropolis weights \cite{b24}) satisfying $\pi ^{s,j} \ge 0$ and $\sum_{j\in\mathcal G^s} \pi ^{s,j} = 1$, $\forall s,j \in \mathcal G$ to ensure that the consensus matrix $\mathbf \Pi $ is primitive and doubly stochastic \footnote[1]{A non-negative consensus matrix $\mathbf \Pi $ is doubly stochastic if all its rows and columns sum up to $1$. Further, it is primitive if there exists an integer $l$ such that all the elements of $\mathbf \Pi{}^l$ are strictly positive \cite{b31}.}, we have $\lim _{l \to \infty } \pi^{s,j}_l = \frac{1}{\lvert \mathcal G \rvert}$ \cite{b31}. Here, $\pi ^{s,j}$ is the $(s,j)$-th entry of the  matrix $\mathbf \Pi $, $\pi^{s,j}_l$ denotes the $(s,j)$-th entry of $\mathbf \Pi{}^l$, and $\lvert \mathcal G \rvert$ is the cardinality of $\mathcal G$. Then, for a state set $\{ \bm x_s \}_{s \in \mathcal G}$, the state $\bm x_s$ on node $s$ is updated at iteration $l$ as $\bm x_s (l) = \sum_{j\in\mathcal G^s} \pi ^{s,j} \bm x_s (l-1)$ with initialization $\bm x_s(0) = \bm x_s$, resulting in $\lim _{l \to \infty } \bm x_s (l) = \frac{1}{\lvert \mathcal G \rvert} \sum_{s\in\mathcal G} \bm x_s $, $\forall s \in \mathcal G$.
\end{definition}

Assume that, at time $k$, each node $s \in \mathcal G$ provides an information set $\{\hat{\bm q}_{k,s}^{x[i]}, \hat{\bm q}_{k,s}^{p[i]}, \mathbf{\Omega}_{k,s}^{x[i]}, \mathbf{\Omega}_{k,s}^{p[i]}\}$. Then, at each iteration, each node $s$ uses CC fusion to calculate a linear combination of quantities in $\mathcal G^s$ with suitable weights $\pi^{s,j}$, $j\in \mathcal G^s$. By alternatively doing this between nodes, the iteration operation yields the averages $\sum_{s \in \mathcal{G}}\{\hat{\bm q}_{k,s}^{x[i]}, \hat{\bm q}_{k,s}^{p[i]}, \mathbf{\Omega}_{k,s}^{x[i]}, \mathbf{\Omega}_{k,s}^{p[i]}\} /\lvert \mathcal G \rvert$ when the number of iterations $L$ approaches to $\infty$. Thus, the ultimate outputs are $\hat{\bm x}_{k,s} = \sum_{s \in \mathcal{G}} \mathbf{\Omega}_{k,s}^{x[n_k]} \backslash \sum_{s \in \mathcal{G}}(\mathbf{\Omega}_{k,s}^{x[n_k]} \hat{\bm x}_{k,s}^{x[n_k]})$ and $\hat{\bm p}_{k,s} = \sum_{s \in \mathcal{G}} \mathbf{\Omega}_{k,s}^{p[n_k]} \backslash  \sum_{s \in \mathcal{G}}(\mathbf{\Omega}_{k,s}^{p[n_k]} \hat{\bm p}_{k,s}^{x[n_k]})$. In fact, $L$ linearly increases both computation and communication burdens, and thus the maximum $L$ should be a trade-off between cost and performance. CI filter consists of the following three steps: \\ 
(1) Measurement Update Step (Correction)\\
If node $s \in \mathcal S$, compute
\begin{subequations} \label{eq22}
	\begin{equation} \label{eq22a}
		\hat{\bm q}_{k,s}^{x[i]} = \hat{\bm q}_{k,s}^{x[i-1]}+ \mathbf H ^{\mathsf{T}} \mathbf V_{k,s}^{x[i]} \bm{y}_{k,s}^{i}
	\end{equation}
	\begin{equation}\label{eq22b}
		\mathbf{\Omega}_{k,s}^{x[i]} = \mathbf{\Omega}_{k,s}^{x[i-1]}+\mathbf H ^{\mathsf{T}} \mathbf V_{k,s}^{x[i]} \mathbf{H} 
	\end{equation}
\end{subequations}

\begin{subequations} \label{eq23}
	\begin{equation} \label{eq23a}
		\hat{\bm q}_{k,s}^{p[i]}=\hat{\bm q}_{k,s}^{p[i-1]}+ \left( \hat{\mathbf M}_{k,s}^{[i]}\right)^{\mathsf{T}} \mathbf V_{k,s}^{p[i]} \tilde{\mathbf Y}_{k,s}^{[i]}
	\end{equation}
	\begin{equation} \label{eq23b}
		\mathbf{\Omega}_{k,s}^{p[i]} = \mathbf{\Omega}_{k,s}^{p[i-1]}+\left( \hat{\mathbf M}_{k,s}^{[i]}\right)^{\mathsf{T}} \mathbf V_{k,s}^{p[i]} \hat{\mathbf M}_{k,s}^{[i]}.
	\end{equation}
\end{subequations}
If node $s \in \mathcal C$, let
\begin{subequations} \label{eq24}
	\begin{equation} \label{eq24a}
		\hat{\bm q}_{k,s}^{x[i]} = \hat{\bm q}_{k,s}^{x[i-1]},\:\mathbf{\Omega}_{k,s}^{x[i]} = \mathbf{\Omega}_{k,s}^{x[i-1]}
	\end{equation}
	\begin{equation} \label{eq24b}
		\hat{\bm q}_{k,s}^{p[i]}=\hat{\bm q}_{k,s}^{p[i-1]},\:\mathbf{\Omega}_{k,s}^{p[i]} = \mathbf{\Omega}_{k,s}^{p[i-1]}.
	\end{equation}
\end{subequations} 
(2) Consensus Step (Consensus)\\
Perform CC fusion on $\hat{\bm q}_{k,s}^{x[i]}, \hat{\bm q}_{k,s}^{p[i]}, \mathbf{\Omega}_{k,s}^{x[i]}, \mathbf{\Omega}_{k,s}^{p[i]}$ independently for $L$ iterations ($L$ is designed a priori).\\
(3) Time Update Step (Prediction) \\
Perform \eqref{eq20} and \eqref{eq21} to accomplish the prediction step.\\

\begin{algorithm}[h]
	\caption{Consensus on Information (CI) Filter}\label{algorithm2}	
	\SetAlgoLined
	\textbf{Initialization:} $\hat{\bm x}_{1,s}^{[0]}$, $\hat{\bm p}_{1,s}^{[0]}$, $\mathbf{\Omega}_{1,s}^{x[0]}$, and $\mathbf{\Omega}_{1,s}^{p[0]}$ \;
	\For{$k \gets 1,2,\cdots $ \tcp*[h]{scan time} }{
		\textbf{Data:} $\{\bm y_{k,s}^{i} \}^{n_{k}}_{i=1} (s \in \mathcal N)$ \;
		Initialization: $\hat{\bm x}_{k,s}^{[0]}$, $\hat{\bm p}_{k,s}^{[0]}$, $\mathbf{\Omega}_{k,s}^{x[0]}$, and $\mathbf{\Omega}_{k,s}^{p[0]}$ \;
		
		\For{$i=1,2,\cdots,n_k$ \tcp*[h]{sequential} }{\textbf{Correction} \; 
			\eIf{$s \in \mathcal S$}{compute $\hat{\bm q}_{k,s}^{x[i]},\mathbf{\Omega}_{k,s}^{x[i]},\hat{\bm q}_{k,s}^{p[i]},\mathbf{\Omega}_{k,s}^{p[i]}$ via \eqref{eq22} and \eqref{eq23}}{compute $\hat{\bm q}_{k,s}^{x[i]},\mathbf{\Omega}_{k,s}^{x[i]},\hat{\bm q}_{k,s}^{p[i]},\mathbf{\Omega}_{k,s}^{p[i]}$ via \eqref{eq24a} and \eqref{eq24b}}
			\textbf{Consensus operation} \;
			set $\hat{\bm q}_{k,s}^{x[i]}(0)\gets\hat{\bm q}_{k,s}^{x[i]}$ , $\mathbf{\Omega}_{k,s}^{x[i]}(0)\gets\mathbf{\Omega}_{k,s}^{x[i]}$, $\hat{\bm q}_{k,s}^{p[i]}(0)\gets\hat{\bm q}_{k,s}^{p[i]}$, $\mathbf{\Omega}_{k,s}^{p[i]}(0)\gets\mathbf{\Omega}_{k,s}^{p[i]}$	\;
			\For{$l = 0,\cdots,L-1$}{perform AA on $\hat{\bm q}_{k,s}^{x[i]}, \hat{\bm q}_{k,s}^{p[i]}, \mathbf{\Omega}_{k,s}^{x[i]}, \mathbf{\Omega}_{k,s}^{p[i]}$}
			let $\hat{\bm q}_{k,s}^{x[i]} \gets\hat{\bm q}_{k,s}^{x[i]}(L)$, $\mathbf{\Omega}_{k,s}^{x[i]} \gets\mathbf{\Omega}_{k,s}^{x[i]}(L)$ 
			$\hat{\bm q}_{k,s}^{p[i]} \gets\hat{\bm q}_{k,s}^{p[i]}(L)$, $\mathbf{\Omega}_{k,s}^{p[i]}\gets\mathbf{\Omega}_{k,s}^{p[i]}(L)$
		}
		\textbf{Output1:} $\hat{\bm{x}}_{k,s}\gets\left( \mathbf{\Omega}_{k,s}^{x[n_k]}\right)^{-1}\hat{\bm q}_{k,s}^{x[n_k]}$, $\mathbf C_{k,s}^{x} \gets \left( \mathbf{\Omega}_{k,s}^{x[n_k]}\right)^{-1}$ \;
		
		\textbf{Output2:} $\hat{\bm{p}}_{k,s}\gets\left( \mathbf{\Omega}_{k,s}^{p[n_k]}\right)^{-1}\hat{\bm q}_{k,s}^{p[n_k]}$, $\mathbf C_{k,s}^{p} \gets \left( \mathbf{\Omega}_{k,s}^{p[n_k]}\right)^{-1}$\;
		\textbf{Prediction:} as in Algorithm \ref{algorithm1}	
	}
\end{algorithm}

The detailed CI filter is shown in Algorithm \ref{algorithm2}. It is worth noting that CI filter gives a convergent solution even in a single consensus iteration ($L=1$) \cite{b25}. Moreover, the performance of CI filter is less sensitive to more measurements since it directly broadcasts the information quantities to achieve consensus.
\subsection{Distributed Consensus on Measurement Filter} 
\label{subCM} 
Since the dynamic models for both the kinematics and extent are the same on each node, the local prediction step has an identical form to its centralized counterpart. Thus, the remaining objective in a distributed filter is to compute the summation terms of innovation parts 
\begin{gather*}
	\triangle \hat{\bm q}_{k}^{x[i]}:= \sum_{s \in \mathcal{S}} \mathbf H^{\mathsf{T}} \mathbf V_{k,s}^{x[i]} \bm{y}_{k,s}^{i},\:\triangle \mathbf {\Omega}_{k}^{x[i]}:= \sum_{s\in\mathcal S}\mathbf H^{\mathsf{T}} \mathbf V_{k,s}^{x[i]} \mathbf{H}  \\
	\triangle \hat{\bm q}_{k}^{p[i]}:= \sum_{s \in \mathcal{S}} \left( \hat{\mathbf M}_{k,s}^{[i-1]}\right)^{\mathsf{T}} \mathbf V_{k,s}^{p[i]} \tilde{\mathbf Y}_{k,s}^{[i]}\\
	\triangle \mathbf {\Omega}_{k}^{p[i]}:= \sum_{s \in \mathcal{S}}\left( \hat{\mathbf M}_{k,s}^{[i-1]}\right)^{\mathsf{T}} \mathbf V_{k,s}^{p[i]} \hat{\mathbf M}_{k,s}^{[i-1]}
\end{gather*}
as shown in \eqref{eq18} and \eqref{eq19} via a distributed way \cite{b25}. Once the objective is complete, the consensus is reached with the assistance of measurement-to-measurement fusion. The effective fusion method falls into the CM scope. To this end, each node $s \in \mathcal G$ computes a linear combination of innovation parts from its neighboring nodes $j \in \mathcal G^s$ to update its values by using CC fusion. The operation generates approximately averaged values on each node $s$ (here, the maximum iteration $L$ is a suitable value) when the consensus is complete, while CEOT filter needs $\left\{\triangle \mathbf {\Omega}_{k}^{p[i]}, \triangle \hat{\bm q}_{k}^{p[i]}, \triangle \mathbf {\Omega}_{k}^{x[i]}, \triangle \hat{\bm q}_{k}^{x[i]} \right\}$. This inconsistency is partially compensated by multiplying a wight $\omega_{k,s}$ \cite{b25}. The CM information filter consists of the following four steps: \\
(1) Compute local innovation parts \\
If node $s \in \mathcal S$, let
\begin{subequations} \label{eq25}
	\begin{equation} \label{eq25a}
		\delta\hat{\bm q}_{k,s}^{x[i]}= \mathbf H^{\mathsf{T}} \mathbf V_{k,s}^{x[i]} \bm{y}_{k,s}^{i}, \: \delta\mathbf{\Omega}_{k,s}^{x[i]}=\mathbf H^{\mathsf{T}} \mathbf V_{k,s}^{x[i]} \mathbf{H} 
	\end{equation}
	\begin{equation} \label{eq25b}
		\delta\hat{\bm q}_{k,s}^{p[i]}=\left( \hat{\mathbf M}_{k,s}^{[i-1]}\right)^{\mathsf{T}} \mathbf V_{k,s}^{p[i]} \tilde{\mathbf Y}_{k,s}^{[i]},  
	\end{equation}
	\begin{equation}\label{eq25c}
		\delta\mathbf{\Omega}_{k,s}^{p[i]}=\left( \hat{\mathbf M}_{k,s}^{[i-1]}\right)^{\mathsf{T}} \mathbf V_{k,s}^{p[i]} \hat{\mathbf M}_{k,s}^{[i-1]}.
	\end{equation}
\end{subequations}  
If node $s \in \mathcal C$, let
\begin{equation} \label{eq26}
	\delta\hat{\bm q}_{k,s}^{x[i]}=\bm 0,\: \delta\mathbf{\Omega}_{k,s}^{x[i]}=\bm 0 ,\:\delta\hat{\bm q}_{k,s}^{p[i]}=\bm 0,\:\delta\mathbf{\Omega}_{k,s}^{p[i]}=\bm 0.
\end{equation}
(2) Consensus Step (Consensus)\\
Perform CC fusion on $\delta\mathbf{\Omega}_{k,s}^{x[i]}, \delta\hat{\bm q}_{k,s}^{x[i]}, \delta\mathbf{\Omega}_{k,s}^{p[i]}, \delta\hat{\bm q}_{k,s}^{p[i]}$ independently for $L$ iterations ($L$ is designed a priori).\\
(3) Measurement Update Step (Correction)\\
\begin{subequations} \label{eq27}
	\begin{equation} \label{eq27a}
		\hat{\bm q}_{k,s}^{x[i]}=\hat{\bm q}_{k,s}^{x[i-1]}+ \omega_{k,s}\delta\hat{\bm q}_{k,s}^{x[i]}(L)
	\end{equation}
	\begin{equation} \label{eq27b}
		\mathbf{\Omega}_{k,s}^{x[i]}=\mathbf{\Omega}_{k,s}^{x[i-1]}+\omega_{k,s} \delta \mathbf{\Omega}_{k,s}^{x[i]}(L)
	\end{equation}
\end{subequations}

\begin{subequations} \label{eq28}
	\begin{equation} \label{eq28a}
		\hat{\bm q}_{k,s}^{p[i]}=\hat{\bm q}_{k,s}^{p[i-1]}+ \omega_{k,s} \delta\hat{\bm q}_{k,s}^{p[i]}(L)
	\end{equation}
	\begin{equation} \label{eq28b}
		\mathbf{\Omega}_{k,s}^{p[i]}=\mathbf{\Omega}_{k,s}^{p[i-1]}+\omega_{k,s}\delta\mathbf{\Omega}_{k,s}^{p[i]}(L).
	\end{equation}
\end{subequations}
(4) Time Update Step (Prediction) \\
The prediction step is the same as shown in CI filter.

\begin{algorithm}[h]
	\caption{Consensus on Measurement (CM) Filter}\label{algorithm3}	
	\SetAlgoLined
	\textbf{Initialization:} $\hat{\bm x}_{1,s}^{[0]}$, $\hat{\bm p}_{1,s}^{[0]}$, $\mathbf{\Omega}_{1,s}^{x[0]}$, and $\mathbf{\Omega}_{1,s}^{p[0]}$ \;
	\For{$k \gets 1,2,\cdots $ \tcp*[h]{scan time} }{
		\textbf{Data:} $\{\bm y_{k,s}^{i} \}^{n_{k}}_{i=1} (s \in \mathcal N)$ \;
		Initialization: $\hat{\bm x}_{k,s}^{[0]}$, $\hat{\bm p}_{k,s}^{[0]}$, $\mathbf{\Omega}_{k,s}^{x[0]}$, and $\mathbf{\Omega}_{k,s}^{p[0]}$ \;
		
		\For{$i=1,2,\cdots,n_k$ \tcp*[h]{sequential} }{\textbf{Compute local innovation parts} \; 
			\eIf{$s \in \mathcal S$}{compute $\delta\hat{\bm q}_{k,s}^{x[i]},\delta\mathbf{\Omega}_{k,s}^{x[i]},\delta\hat{\bm q}_{k,s}^{p[i]},\delta\mathbf{\Omega}_{k,s}^{p[i]}$ via \eqref{eq25}}{compute $\delta\hat{\bm q}_{k,s}^{x[i]},\delta\mathbf{\Omega}_{k,s}^{x[i]},\delta\hat{\bm q}_{k,s}^{p[i]},\delta\mathbf{\Omega}_{k,s}^{p[i]}$ via \eqref{eq26}}
			\textbf{Consensus operation} \;
			set $\delta \hat{\bm x}^{[i]}_{k,s}(0) \gets\delta \hat{\bm x}^{[i]}_{k,s}$ , $\delta \mathbf{\Omega}^{x[i]}_{k,s}(0) \gets \delta \mathbf{\Omega}^{x[i]}_{k,s}$,  $\delta \hat{\bm p}^{[i]}_{k,s}(0)\gets\delta \hat{\bm p}^{[i]}_{k,s}$, $\delta \mathbf{\Omega}^{p[i]}_{k,s}(0) \gets\delta \mathbf{\Omega}^{p[i]}_{k,s}$ 	\;
			\For{$l = 0,\cdots,L-1$}{perform AA on $\delta \hat{\bm x}^{[i]}_{k,s}, \delta \mathbf{\Omega}^{x[i]}_{k,s}, \delta \hat{\bm p}^{[i]}_{k,s}, \delta \mathbf{\Omega}^{p[i]}_{k,s}$}
			\textbf{Correction} \;
			compute $\hat{\bm q}_{k,s}^{x[i]},\mathbf{\Omega}_{k,s}^{x[i]},\hat{\bm q}_{k,s}^{p[i]},\mathbf{\Omega}_{k,s}^{p[i]}$ via \eqref{eq27} and \eqref{eq28}
		}
		\textbf{Output1:} $\hat{\bm{x}}_{k,s}=\left( \mathbf{\Omega}_{k,s}^{x[n_k]}\right)^{-1}\hat{\bm q}_{k,s}^{x[n_k]}$, $\mathbf C_{k,s}^{x} = \left( \mathbf{\Omega}_{k,s}^{x[n_k]}\right)^{-1}$ \;
		\textbf{Output2:} $\hat{\bm{p}}_{k,s}=\left( \mathbf{\Omega}_{k,s}^{p[n_k]}\right)^{-1}\hat{\bm q}_{k,s}^{p[n_k]}$, $\mathbf C_{k,s}^{p} = \left( \mathbf{\Omega}_{k,s}^{p[n_k]}\right)^{-1}$ \;
		\textbf{Prediction:} as in Algorithm \ref{algorithm1}	
	}
\end{algorithm}
\section{Stability Analysis}
\label{sec:analysis}
In this section, the stability property of CM filter is analyzed in a linear setting. As for CI filter, one could combine the stability analysis provided in \cite{neuro} with the discussion in CM filter to give the same result. To this end, consider the measurement model \eqref{eq6} with a single measurement (here, the superscript ${i}$ is omitted, and thus the notation $(\cdot)_{k|k-1}$ denotes the corresponding predicted estimates at time $k$) and dynamic model \eqref{eq4}. Meanwhile, we introduce the compensation instrumental diagonal matrix $\bm{\beta}_{k,s} = \mathrm{diag}(\beta_{k,s}^1,\beta_{k,s}^2)$ to eliminate possible approximation error \cite{b24}. Then, \eqref{eq6} is rewritten as follows
\begin{equation} \label{eq29}
	\bm y_{k,s} = \bm{\beta}_{k,s} \mathbf H \bm x_k + \bm v_{k,s}^{x}.
\end{equation}

To verify the boundedness of estimation errors in the mean square for CM filter, the following assumptions are necessary:

\begin{enumerate} 
	\item[A1.] There exists real scalars $\overline{\bm{\beta}}, \overline{\bm f}, \overline{\bm h} \neq 0$ and $\underline{\bm{\beta}}, \underline{\bm f}, \underline{\bm h} \neq 0$ such that for each $k \ge 0$
	\begin{equation}\label{eq30}
		\underline{\bm{\beta}}^2 \mathbf I \le \bm{\beta}_{k,s} \bm{\beta}_{k,s}^{\mathsf{T}} \le \overline{\bm{\beta}}^2 \mathbf I	\:,		\underline{\bm{f}}^2 \mathbf I \le \mathbf F_k^x \mathbf (F_k^x)^{\mathsf{T}} \le \overline{\bm{f}}^2 \mathbf I \:, \underline{\bm{h}}^2 \mathbf I \le \mathbf H \mathbf H^{\mathsf{T}} \le \overline{\bm{h}}^2 \mathbf I
	\end{equation}
	\item[A2.] There exists real scalars $0< {\tau}_{\mathrm{min}} \le {\tau}_{\mathrm{max}}$, $0 < \underline{\omega} \le \overline{\omega}$,  $0 < \underline{r} \le \overline{r}$,  $0 < \underline{q} \le \overline{q}$, $0 < \underline{p} \le \overline{p}$ such that
	\begin{equation} \label{eq31}
		{\tau}_{\mathrm{min}} \le {\tau}^{s} \le {\tau}_{\mathrm{max}}, \: \underline{\omega} \le \omega_{k,s} \le \overline{\omega}, \: \underline{q} \mathbf I \le \mathbf Q_k \le \overline{q} \mathbf I, \: \underline{r} \mathbf I \le \mathbf R_{k,s} \le \overline{r} \mathbf I, \: \underline{p} \le \mathbf{\Omega}_{k,s}^{x}  \le \overline{p}
	\end{equation}
	\item[A3.] The consensus matrix $\mathbf \Pi $ is doubly stochastic and primitive.
\end{enumerate}
With the above assumptions, it is ready to give the following result.

\begin{thm} \label{them1}
	Consider the linear stochastic system given in \eqref{eq4} and \eqref{eq29}. If the consensus iteration $L >1$ in CM filter and the prior information matrix $\mathbf{\Omega}_{1|0,s}^x > 0 $ for any node $s \in \mathcal G$, the estimation error $\bm e_{k+1,s} = \bm x_{k+1} - \hat{\bm x}_{k+1|k+1,s}$ is asymptotically bounded in the mean square, i.e., $\mathop{\limsup}_{k \to \infty} \mathbb{E} \left\{\Vert{\bm e_{k+1,s}}\Vert^2 \right\} < +\infty$, for any node $s \in \mathcal G$ and time $k \ge 0$ under the above assumptions.
\end{thm}

\begin{proofof}[Proof of Theorem \ref{them1}.]
	See Appendix. 
\end{proofof}
\begin{rmk}
 	\begin{itemize}
	\item The boundedness of CM filter does not rely on a specific value of compensation instrumental diagonal matrix $\bm{\beta}_{k,s}$.  
	\item The boundedness of CM filter requires enough iterations, as it only exchanges the innovation parts instead of the information vector and information matrix. 
	\item The whole stability of CM filter is guaranteed only when both the kinematics  and extent error are bounded in the mean square. Here, we only give a discussion of the kinematics error $\bm e_{k,s}$ being bounded in the mean square. One can replace the terms $\mathbf F_k^x \gets \mathbf F_k^p$, $(\mathbf{\Omega}_{k,s}^x)^{-1} \gets (\mathbf{\Omega}_{k,s}^p)^{-1}$, $\mathbf H  \gets \mathbf M_{k,j}$, $\mathbf V_{k,j}^{x} \gets \mathbf V_{k,j}^{p}$, $\bm w_k^x \gets \bm w_k^p$, and ${\bm v}_{k,j}^x \gets ({\bm v}_{k,j}^p - \bar{\bm v}_{k,j}^p)$ in Theorem \ref{them1} to draw the same conclusion about the extent. 
	\end{itemize} 
\end{rmk}
\section{Simulation Examples}
\label{sec:simulation}
In this section, we first evaluate the convergence of sequential processing used in the proposed CEOT, CI and CM. Then, we testify the performance of CEOT, CI, and CM and compare them with a centralized EOT based on the RM model \cite{b19} (abbreviated as CEOT-RM) and a distributed filter in \cite{DEOT} (abbreviated as DEOT). In the end, the consistency of CI and CM is demonstrated in a simple dynamics. All simulations are implemented in  MATLAB--2019b running on a PC (64-bit floating point) with processor \texttt {Intel(R) Core(TM) i7-10510U CPU @ 1.8GHz 2.3GHz and with 20GB RAM.} 

Several standard performance metrics are used in this section: (1) Gaussian Wasserstein/Optimal Sub-Pattern Assignment (OSPA) distance assesses both the position and extent errors \cite{b7,b8}; (2) Computational cost measures the average running time in a filter over all time steps; (3) Normalized estimation error squared (NEES)  checks the consistency; (4) Averaged consensus estimate error (ACEE) evaluates the estimation bias between different nodes (consensus).

The network consists of $14$ communication nodes and $6$ sensor nodes as shown in Fig \ref{link}. The consensus parameter $\pi ^{s,j}$ is computed by the Metropolis weights rule \cite{b24}. As for the scalar weights $\omega_{k,s}$, one can refer to \cite[eq.~(4)]{b25}.
\subsection{Evaluation on the convergence of sequential processing}
In this scenario (S1), we consider a stationary rectangular object to test the convergence of sequential processing used in CI and CM filters. Since the prediction steps are the same among all nodes, we only focus on the correction and consensus steps. The object locates in the origin with lengths $4$ and $9$ meters and it rotates $\frac{\pi}{4}$ from $x$-axis along with the counter-clockwise. True measurements are generated by sensor nodes $s \in \mathcal S$ based on the measurement model \eqref{eq3} with $\mathbf H = \mathbf I_2$. The parameters used in the examined filters are listed in Table \ref{tab1}.

\begin{table}[h]
	\centering
	\caption{Tracker Parameter Setting in S1}\label{tab1}
	\begin{tabular*}{220pt}{cc}
		\hline
		Parameters & Specification \\ \hline
		Meas. Noise Cov. & $\mathbf C_s^v = \mathrm{diag}(3,9)$ \\
		Multi. Noise Cov. & $\mathbf C^h =  \frac{1}{3} \mathbf I_2$ \\
		Pos. estimate & $\hat{\bm x}_{1,s}^{[0]} = [1,1]^{\mathsf{T}}$ \\
		Extent estimate & $\hat{\bm p}_{1,s}^{[0]} = [0,2,12]^{\mathsf{T}}$ \\
		Prior Cov. w.r.t Kine. & $\mathbf {C}_{1,s}^{x[0]} = \mathrm{diag} (1,1)$ \\
		Prior Cov. w.r.t Extent & $\mathbf {C}_{1,s}^{p[0]} = \mathrm{diag} (1, 4, 9)$ \\
		No. of measurement & $100$ \\
		\hline
	\end{tabular*}
\end{table}

\begin{figure}[h]
	\begin{minipage}[hpb]{0.55\linewidth}
		\centering
		\includegraphics[width=3.3in,height=2.1in]{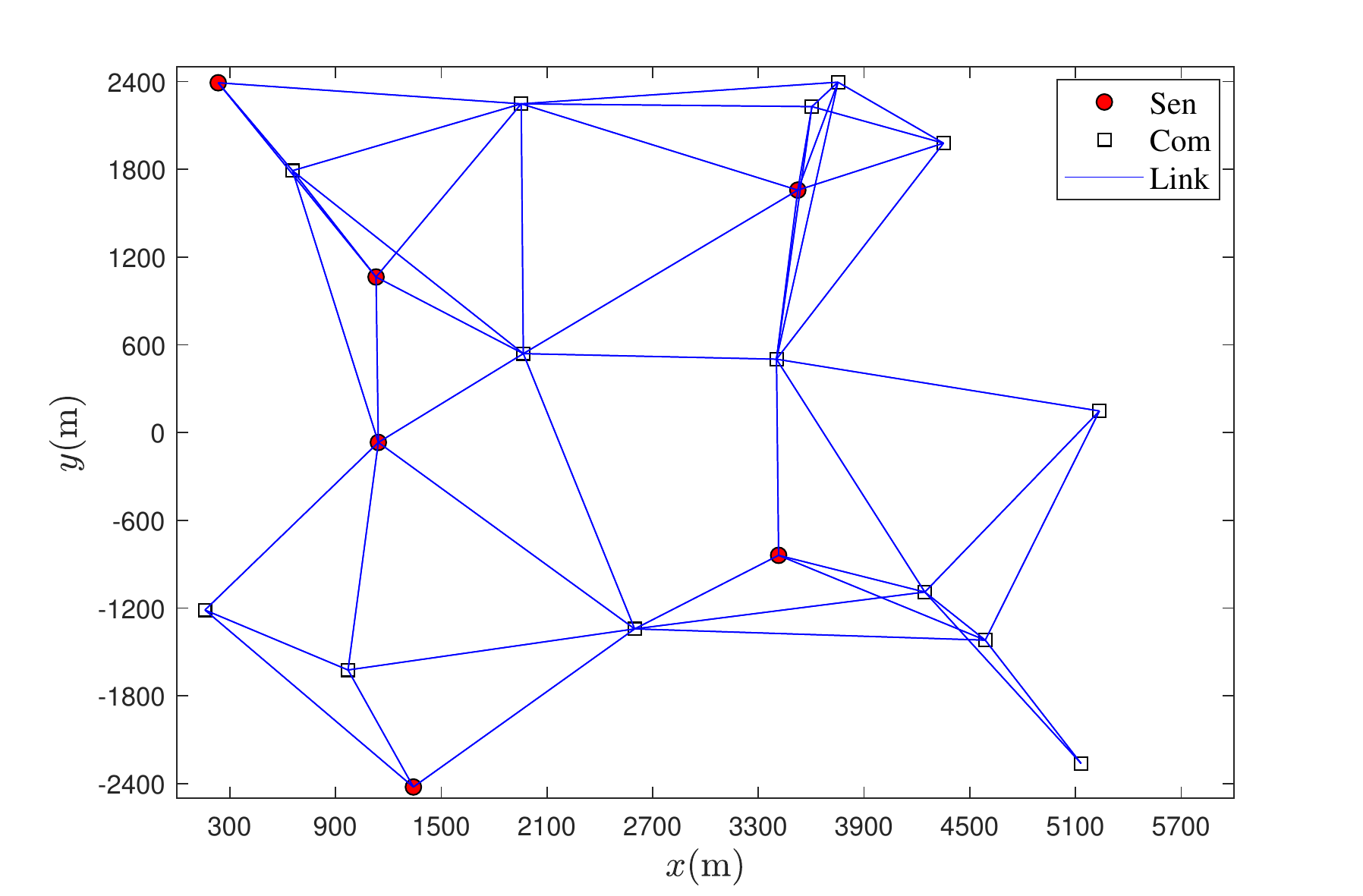}
		\caption{{Sensor network with maximum communication distance $R=2000$m.}}
		\label{link}
	\end{minipage}
	\begin{minipage}[h]{0.45\linewidth}
		\centering
		\includegraphics[width=3.3in,height=2.1in]{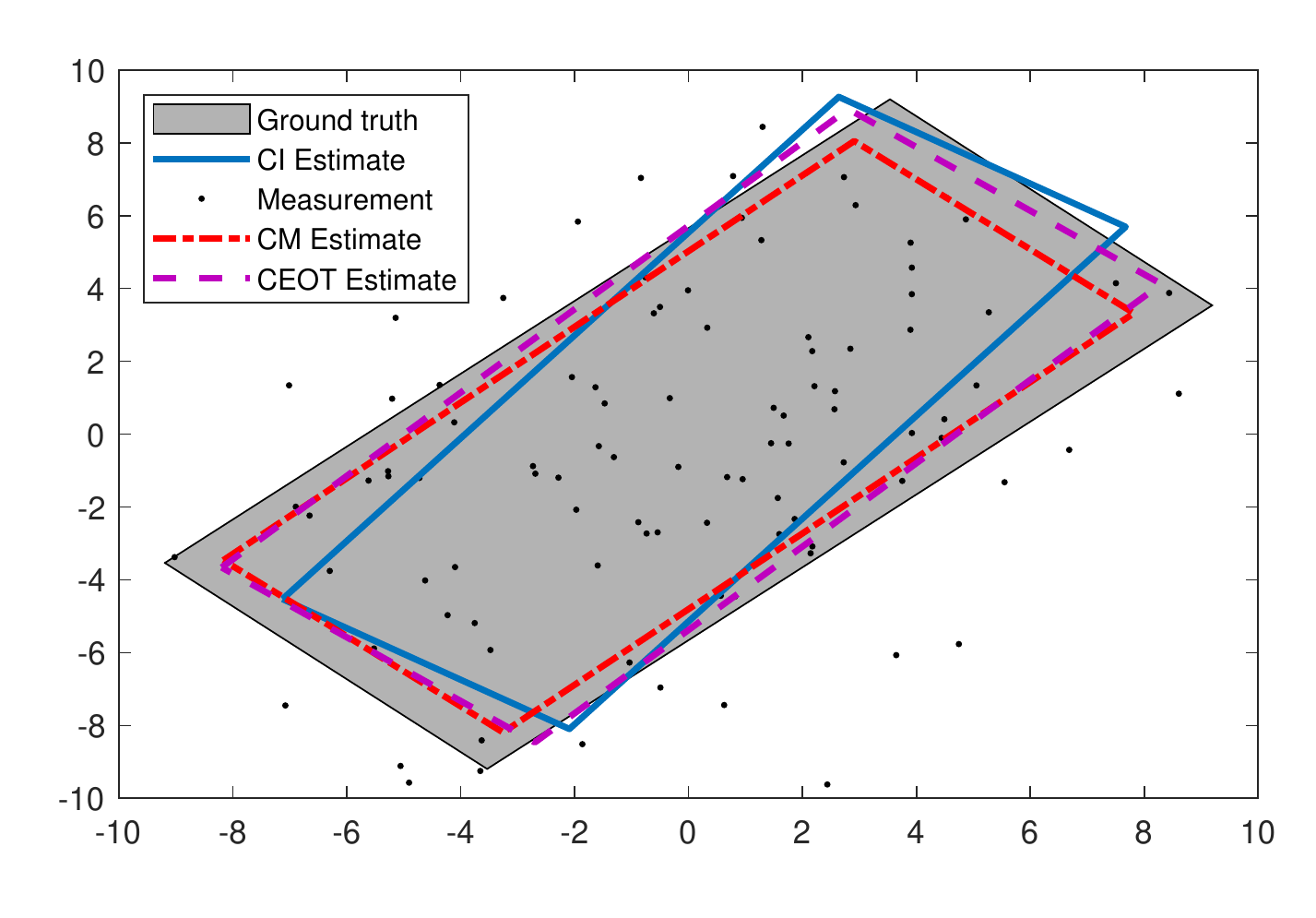}
		\caption{{The results present the ground truth, noisy measurements, and estimates with iteration $L=6$ after $100$ sequential runs.}}
		\label{sta_result}
	\end{minipage}
\end{figure}

\begin{figure}
	\begin{minipage}[thpb]{0.55\linewidth}
		\centering
		\includegraphics[width=3.3in,height=2.1in]{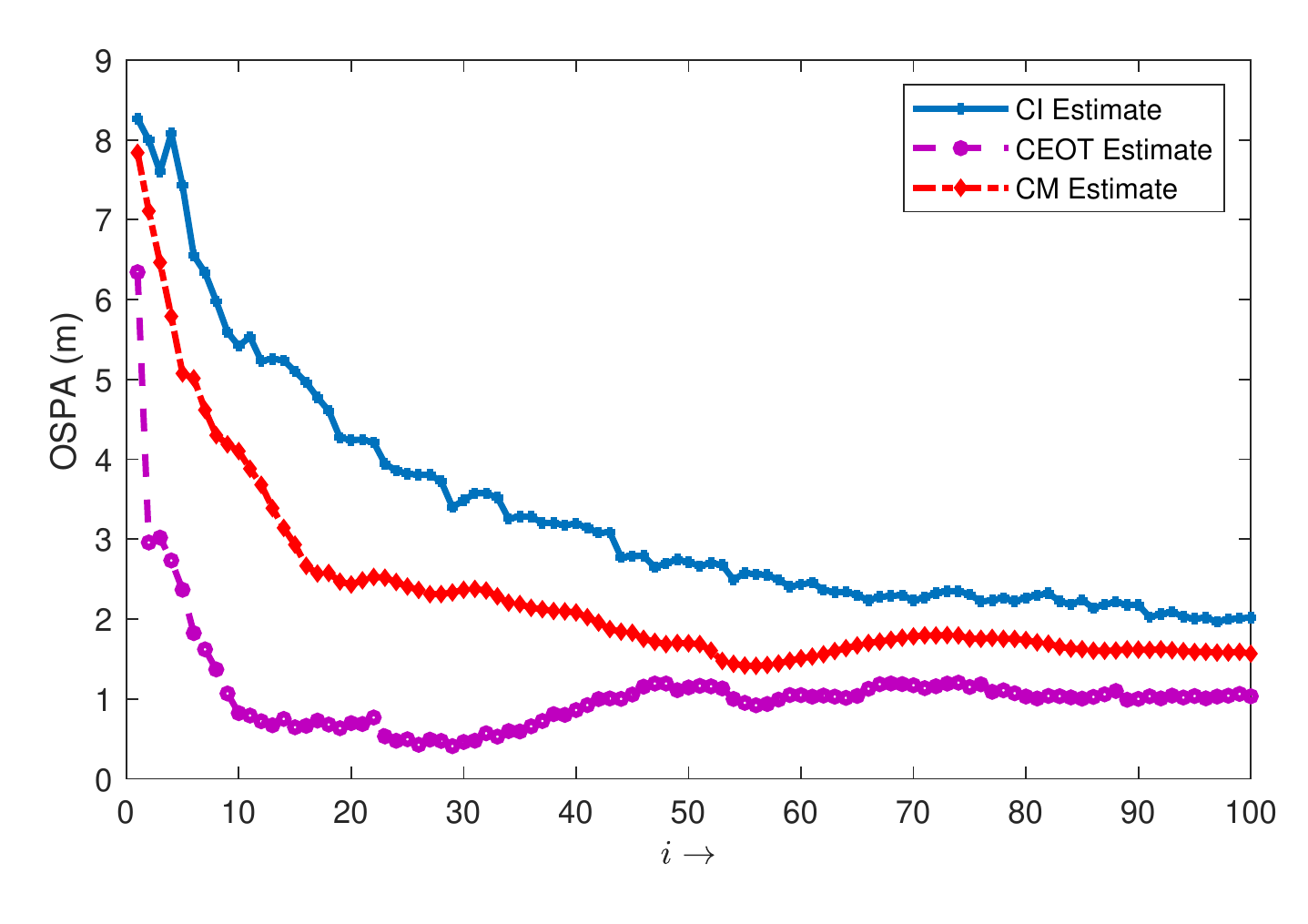}
		\caption{{OSPAs with iteration $L=6$ after $100$ runs.}}
		\label{sta_ospa}
	\end{minipage}
	\begin{minipage}[thpb]{0.45\linewidth}
		\centering
		\includegraphics[width=3.3in,height=2.1in]{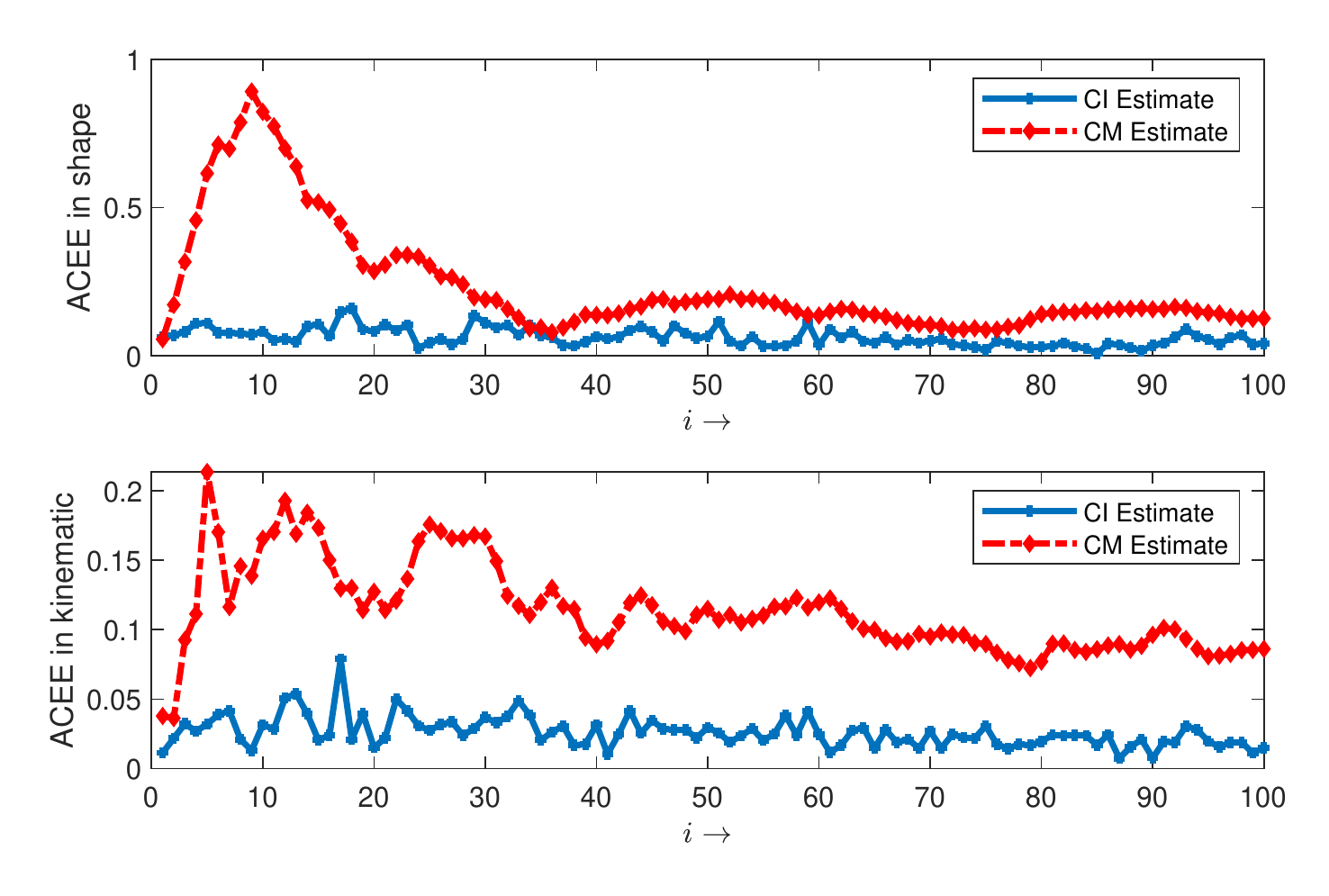}
		\caption{{ACEEs with iteration $L=6$ after $100$ runs.}}
		\label{sta_ACEE}
	\end{minipage}
\end{figure}

The overall tracking results of CM, CI, and CEOT are shown in Fig. \ref{sta_result}. It is shown that CM follows CEOT more tightly than that of CI in this case. To visualize the convergence on both the position and extent, the OSPA distance is used. Here, we select four vertices to uniquely capture the difference of two rectangles in extent. From the OSPAs shown in Fig. \ref{sta_ospa}, some conclusions are drawn: (1) two distributed filters provide a stable and convergent solution after $100$ sequential runs  even in the scenario with high noise intensity; (2) compared with CI, CM shows a faster convergence rate as it broadcasts directly raw measurements instead of estimates across nodes. This exhibits CM's superiority, especially for the high number of measurements.

The ACEE used to check the estimation bias between nodes is defined as 
\begin{equation*}
	\mathrm{ACEE} := \frac{1}{\lvert \mathcal G \rvert (\lvert \mathcal G \rvert-1)} \sum_{s \in \mathcal G} \sum_{j \in \mathcal G} \Vert \hat{\bm x}_{k,s} - \hat{\bm x}_{k,j} \Vert
\end{equation*}
where $\hat{\bm x}_{k,s}$ is the estimate on node $s$. 

Fig. \ref{sta_ACEE} shows that the ACEEs are within a reasonable range for both the kinematics and extent. Here, CI gives lower ACEEs since it uses the information matrix to set the corresponding weights on different nodes. Instead, a weighted mechanism for the estimates is absent in CM.

\subsection{Evaluation on the tracking performance}
\label{sub:7.2}
\subsubsection{Ellipse with Nearly Constant Velocity Model Tracking Scenario}
\label{sub:7.2.1}
\begin{figure}
	\makeatletter\def\@captype{figure}\makeatother
	\begin{minipage}{.45\textwidth}
		\centering
		\includegraphics[scale=.45]{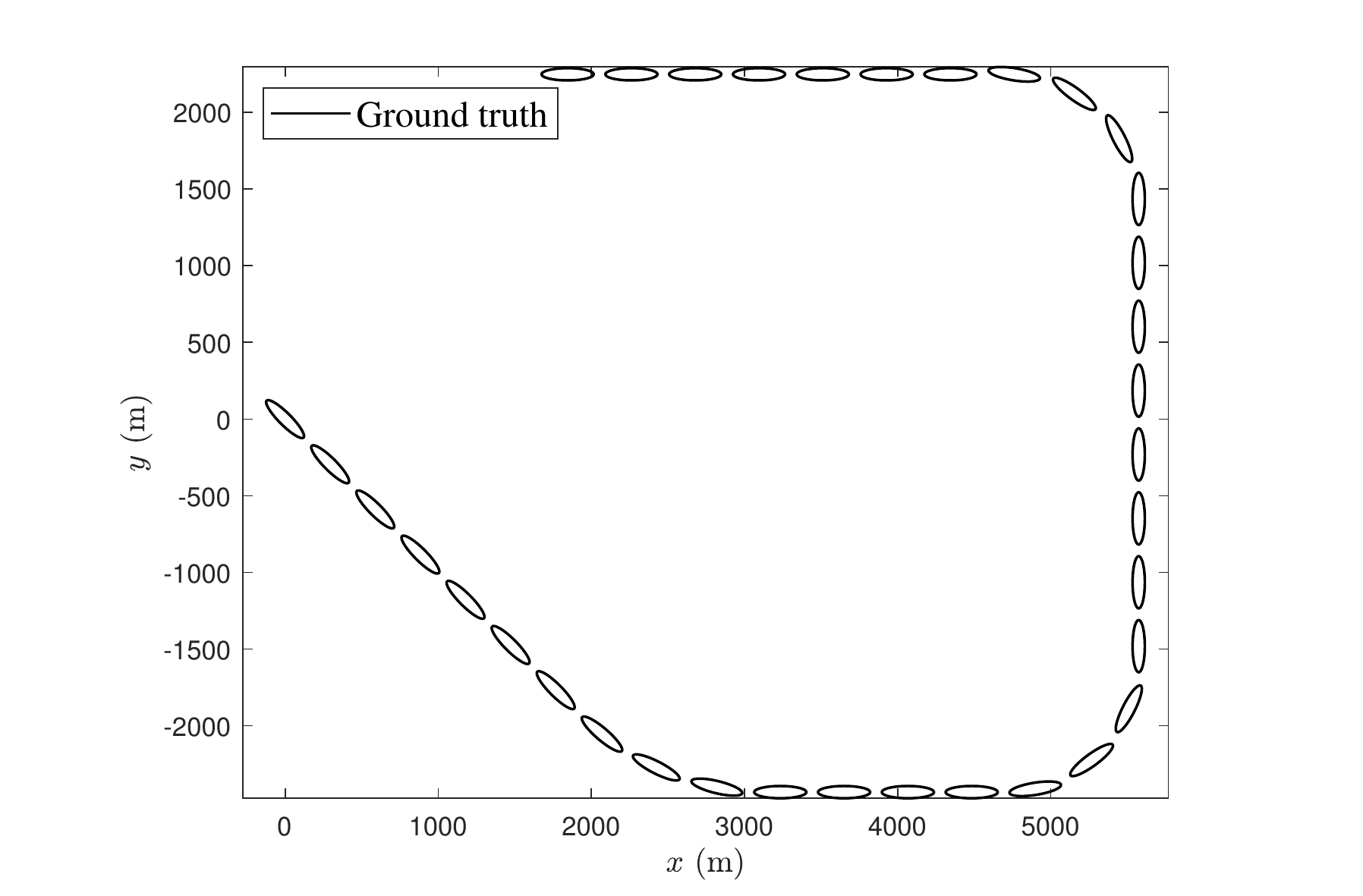}
		\caption{{Trajectory of an elliptical extended object.}} 	\label{tra}
	\end{minipage}
	\makeatletter\def\@captype{table}\makeatother
	\begin{minipage}{.45\textwidth}
		\centering
		\captionof{table}{Tracker Parameter Setting in S2} \label{tab2}	
		\begin{tabular}{cc}
			\hline
			Parameters & Specification \\ \hline
			Scan Time & $\mathrm{T}=10$ s \\
			Meas. Noise Cov. & $\mathbf C_s^v = \mathrm{diag}(200,8)$ \\
			Multi. Noise Cov. & $\mathbf C^h =  \frac{1}{4} \mathbf I_2$ \\
			Cov. in Kine. & $\mathbf C^x_w = \mathrm{diag}(100,100,1,1)$ \\
			Cov. in Extent & $\mathbf C^p_w = \mathrm{diag}(0.05,1,1)$ \\
			Prior Cov. in Kine. & $\mathbf {C}_{1,s}^{x[0]} = \mathrm{diag} (100,100, 10,10)$ \\
			Prior Cov. in Extent & $\mathbf {C}_{1,s}^{p[0]} = \mathrm{diag} (0.36, 70, 40)$ \\
			Degrees of freedom in \cite{b19} & $\alpha_{0|0} = 10$ \\
			Agility constant in \cite{b19} & $\tau = 50$ \\
			\hline
		\end{tabular}
	\end{minipage}
\end{figure}

In this scenario (S2), the object is an ellipse with lengths of the semi-axes $170$m and $40$m. The object moves with nearly constant speed $v=50 \mathrm{km/h}$ following the trajectory as shown in Fig. \ref{tra}. The parameters used in the examined filters are listed in Table \ref{tab2}.
The comparison results focus on the GWD distance, ACEE metric and computation cost over $M=50$ Monte Carlo runs. Moreover, comparison in S2 allows an in-depth analysis of the proposed filters as their performance is varied under different parameters.

\begin{figure}
	\begin{minipage}[thpb]{0.45\linewidth}
		\centering
		\includegraphics[width=3.1in,height=1.8in]{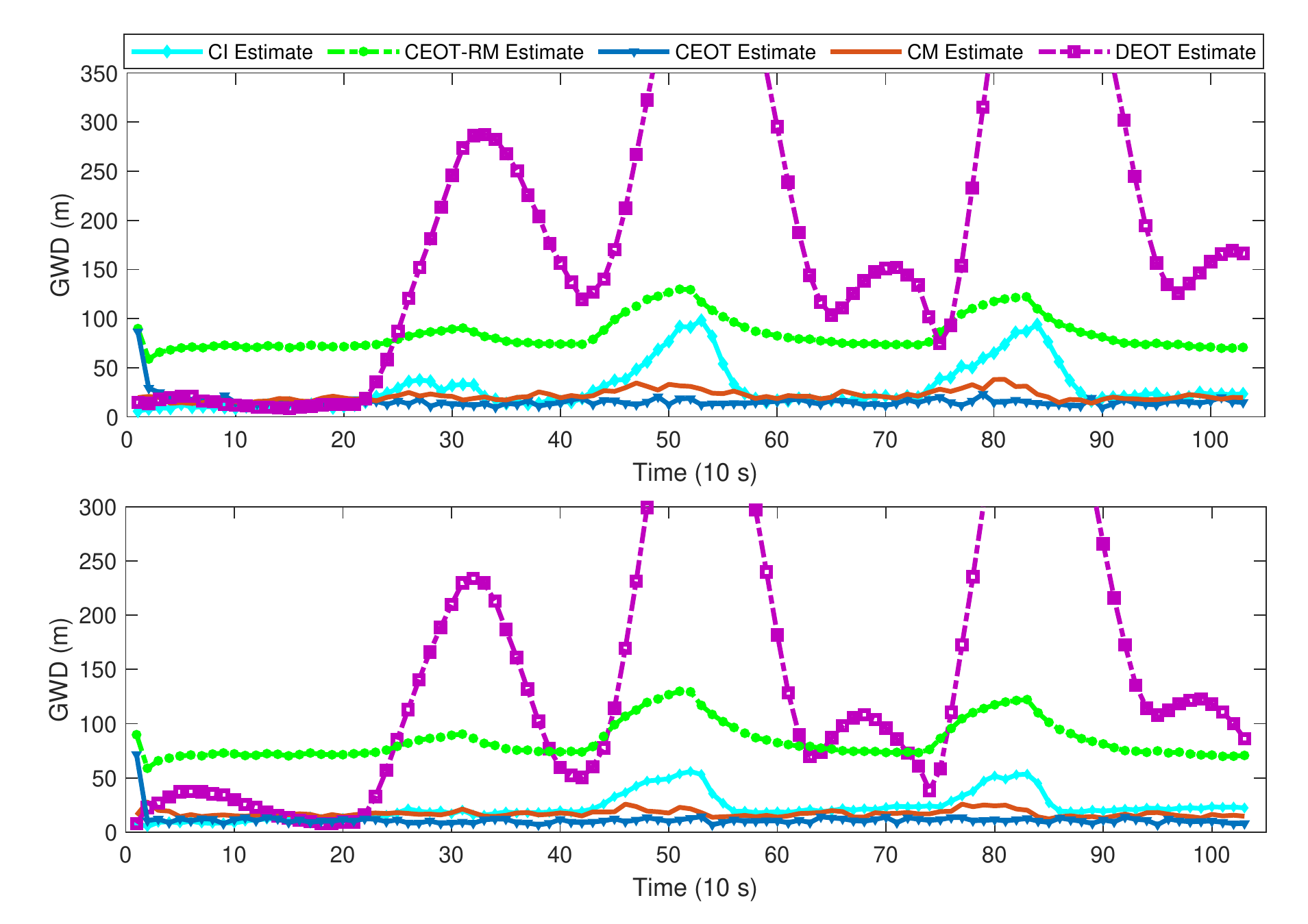}
		\caption{GWDs using CEOT, CEOT-RM, DEOT, CM and CI with iteration $L=6$. The first row shows the GWDs under $\lambda=5$. The second row shows the GWDs under $\lambda=10$.}
		\label{ellip_gwd}
	\end{minipage}
	\begin{minipage}[thpb]{0.55\linewidth}
		\centering
		\includegraphics[width=3.1in,height=1.8in]{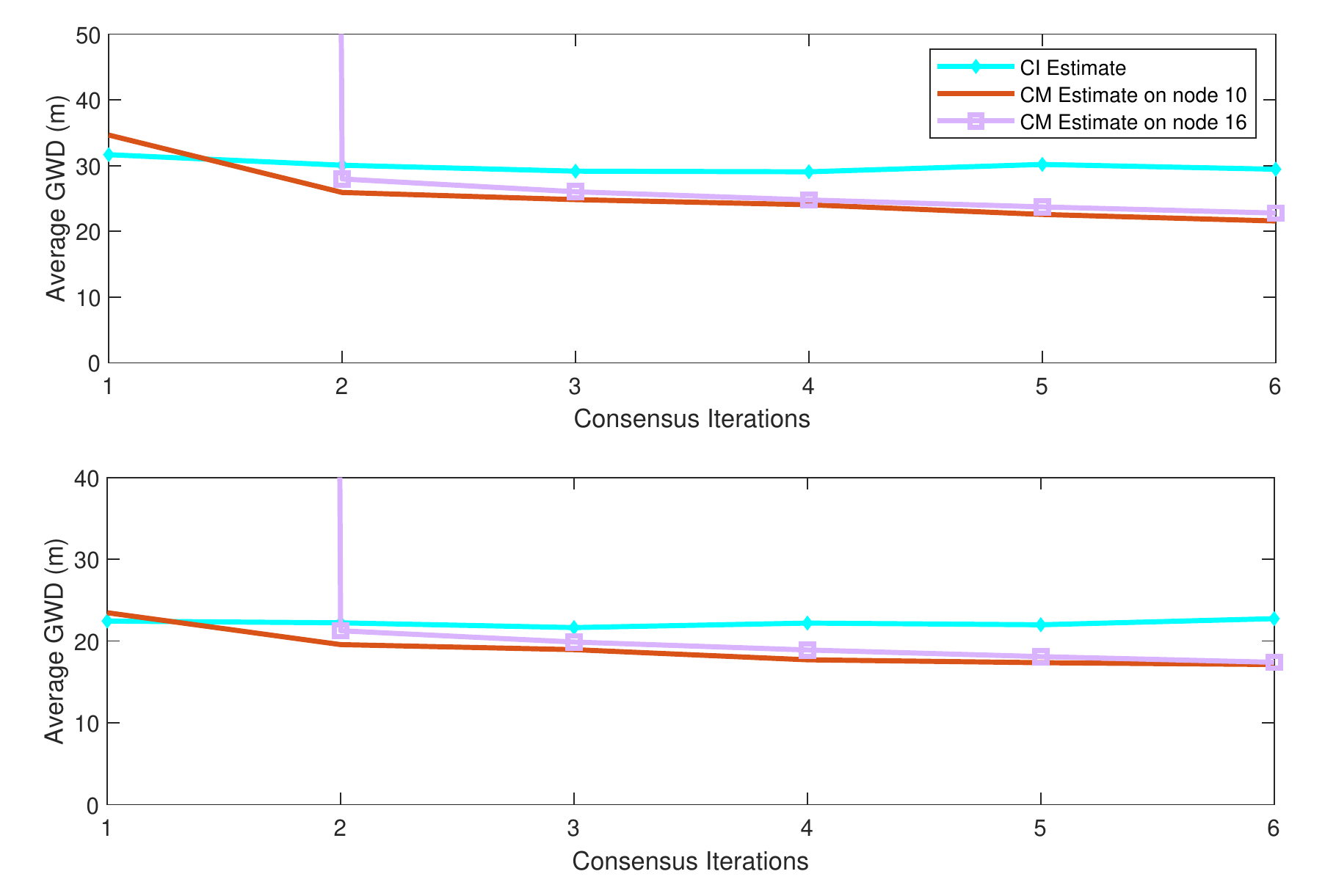}
		\caption{Average GWDs with different iterations. The first row shows the Average GWDs under $\lambda=5$. The second row shows the Average GWDs under $\lambda=10$.}
		\label{ellip_avegwd}
	\end{minipage}
\end{figure}
\begin{figure}
	\begin{minipage}[thpb]{0.45\linewidth}
		\centering
		\includegraphics[width=3.2in,height=2.3in]{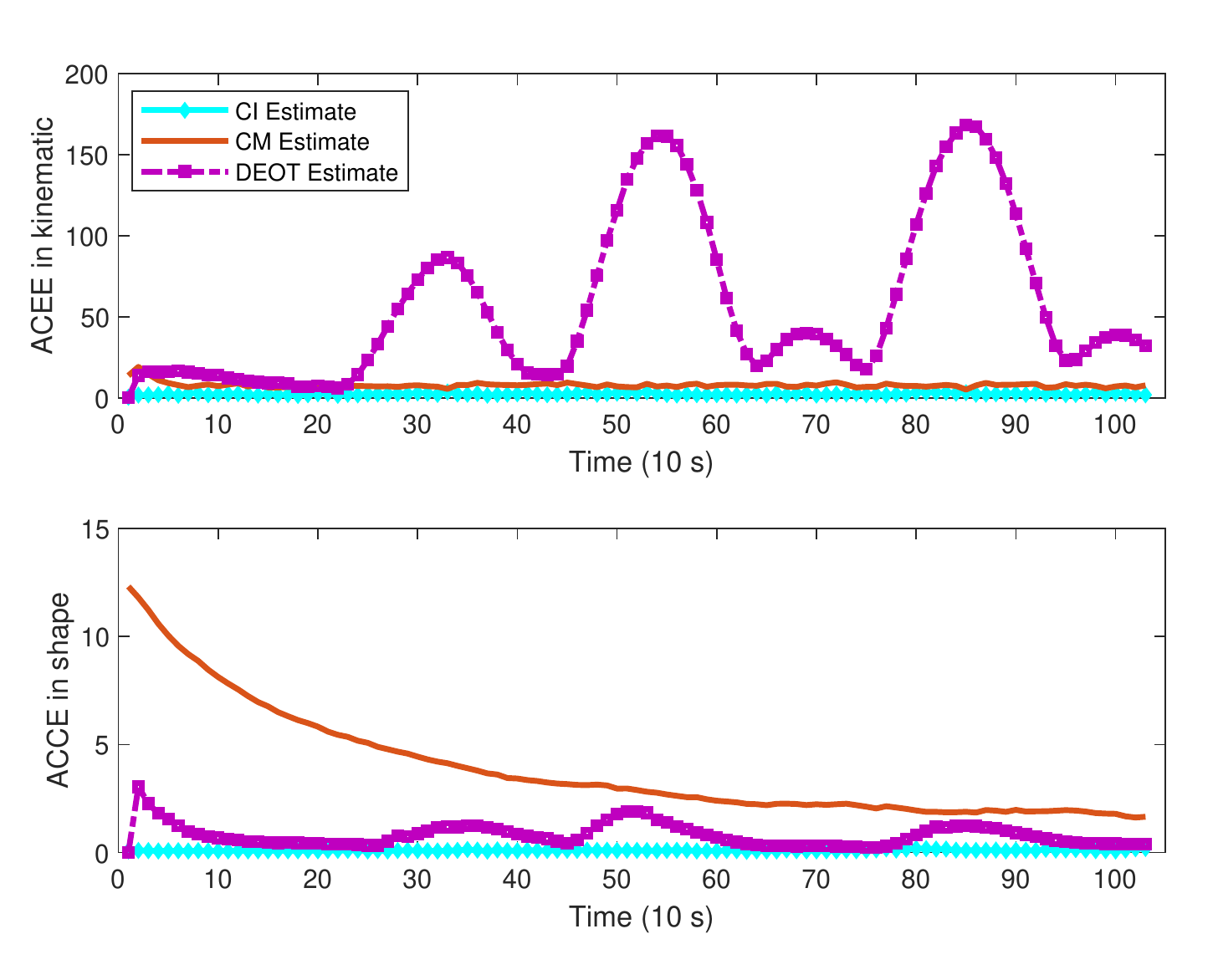}
		\caption{ACEEs using DEOT, CM, and CI with iteration $L=6$ ($\lambda=5$).}
		\label{ellip_acee1}
	\end{minipage}
	\begin{minipage}[thpb]{0.55\linewidth}
		\centering
		\includegraphics[width=3.2in,height=2.3in]{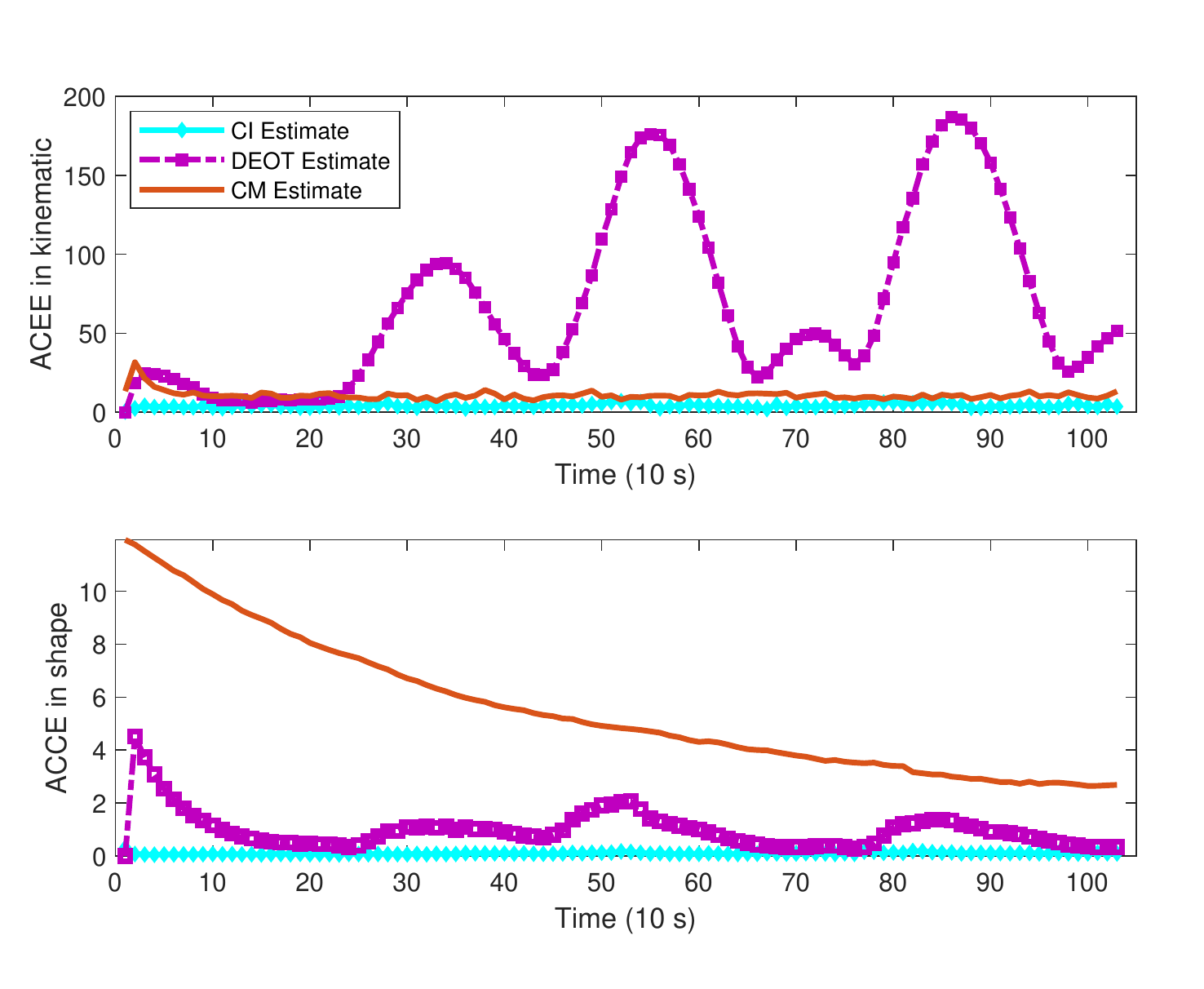}
		\caption{ACEEs using DEOT, CM, and CI with iteration $L=6$ ($\lambda=10$).}
		\label{ellip_acee}
	\end{minipage}
\end{figure}
\begin{figure}
	\begin{minipage}[thpb]{0.45\linewidth}
		\centering
		\includegraphics[width=3.2in,height=2.3in]{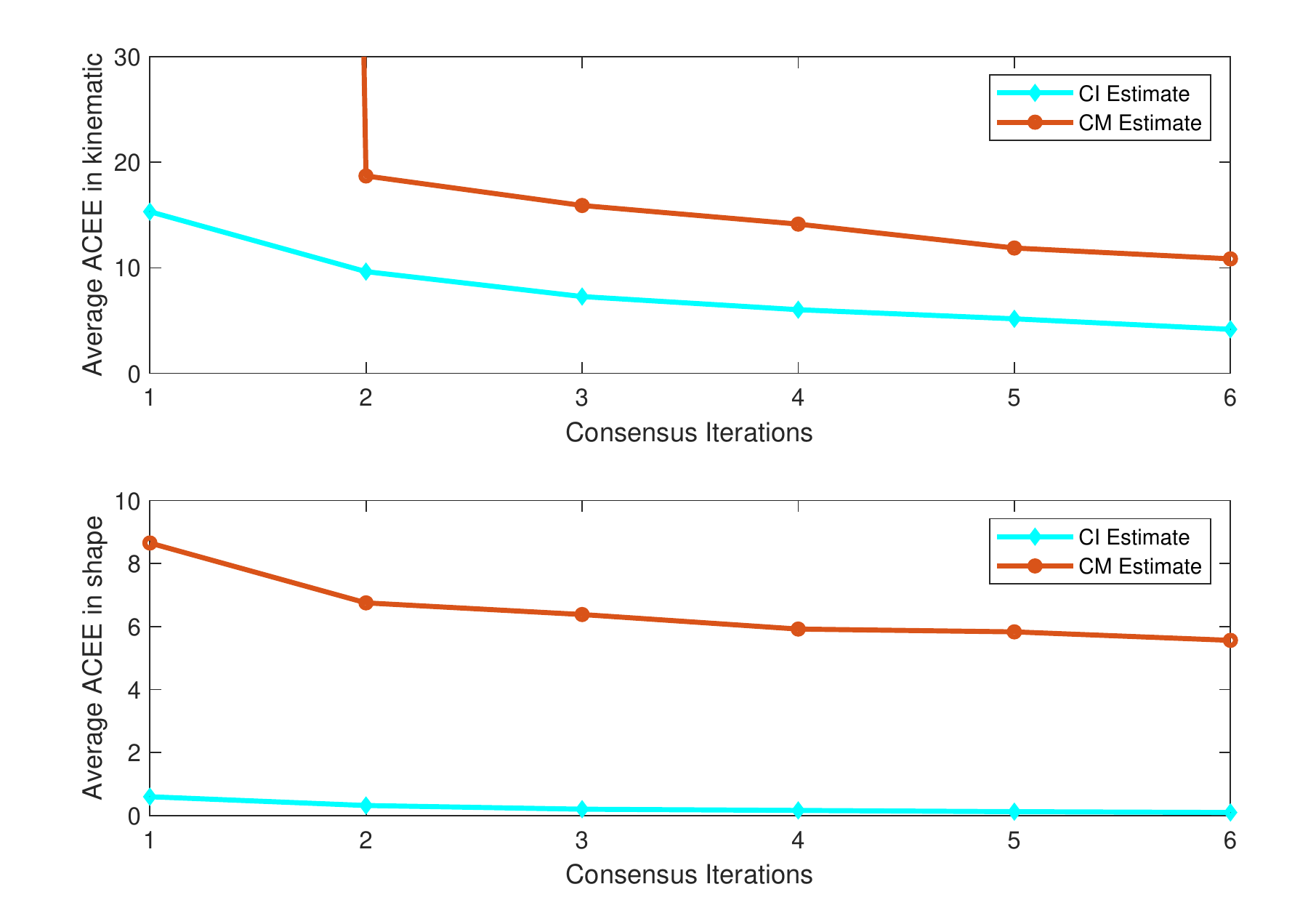}
		\caption{Average ACEEs with different iterations under $\lambda=5$.}
		\label{ellip_aveacee1}
	\end{minipage}
	\begin{minipage}[thpb]{0.55\linewidth}
		\centering
		\includegraphics[width=3.2in,height=2.3in]{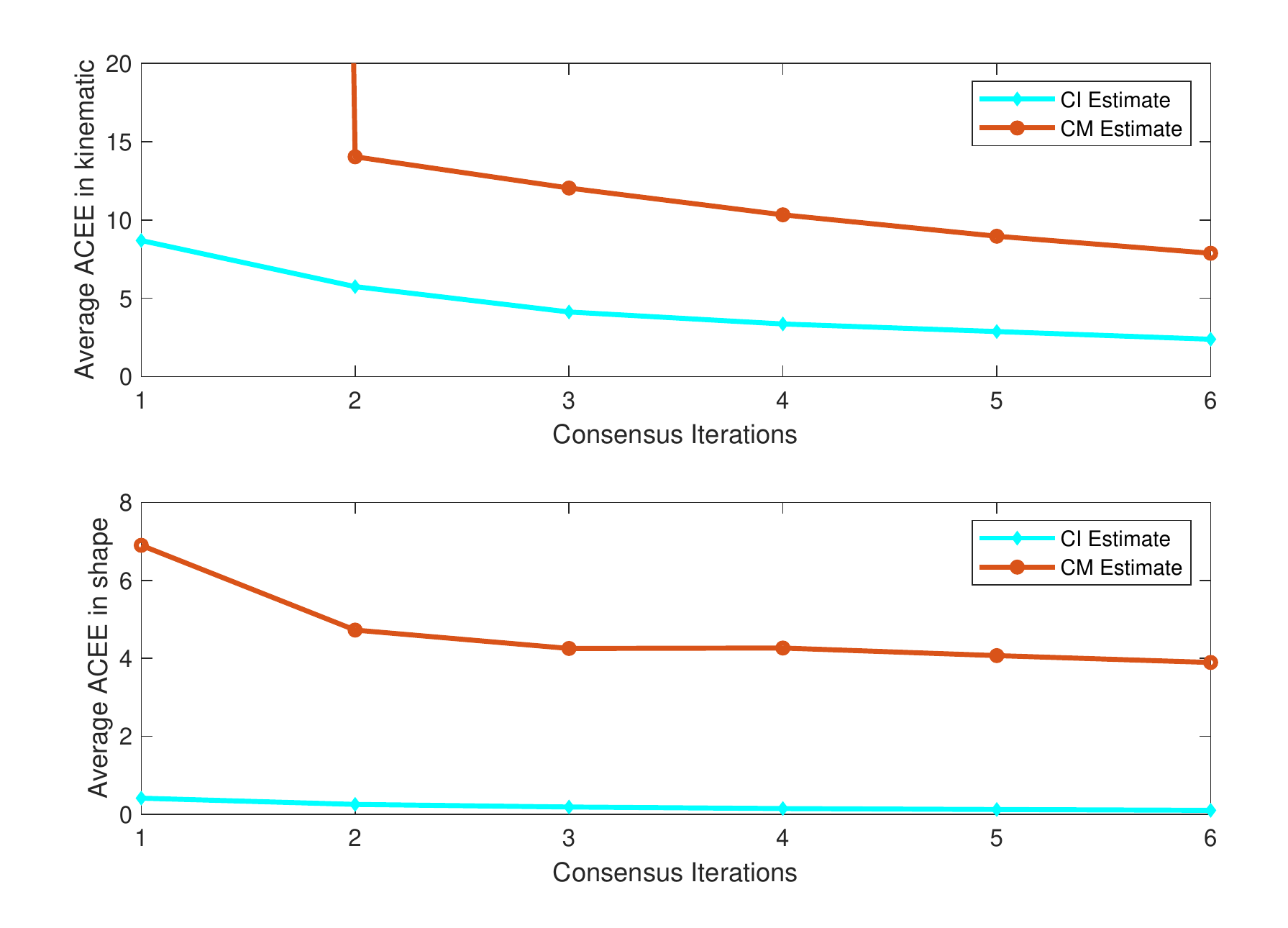}
		\caption{Average ACEEs with different iterations under $\lambda=10$.}
		\label{ellip_aveacee}
	\end{minipage}
\end{figure}
\begin{figure}
	\makeatletter\def\@captype{figure}\makeatother
	\begin{minipage}{.45\textwidth}
		\centering
		\includegraphics[width=3.2in,height=2.3in]{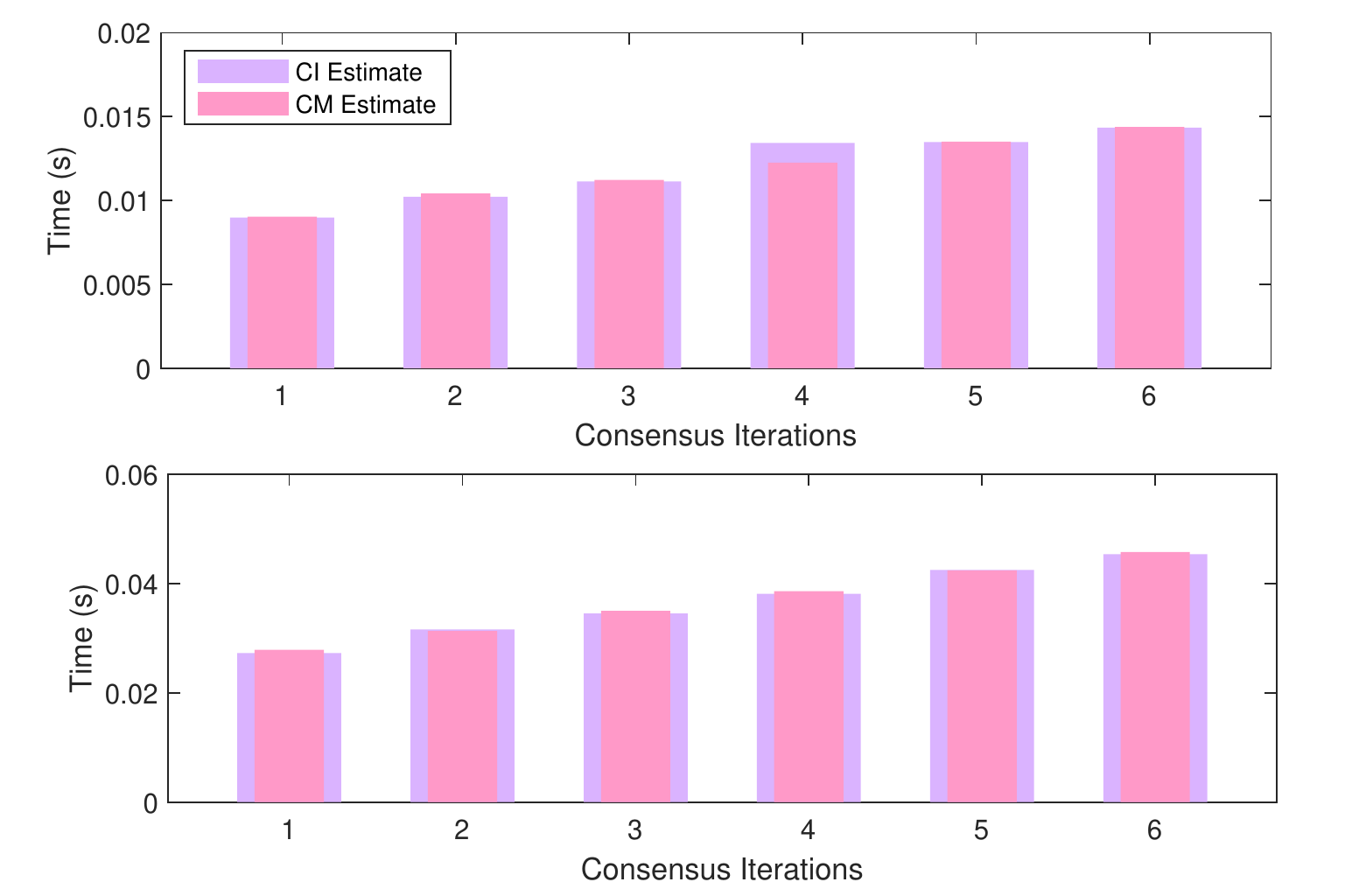}
		\caption{Average computational time (CT) under different iterations. The first row shows the average CT	under $\lambda = 5$, and the second shows the average CT under $\lambda = 10$.}
		\label{ellip_com}
	\end{minipage}
	\makeatletter\def\@captype{table}\makeatother
	\begin{minipage}{.6\textwidth}
		\centering
		\caption{Tracker Parameter Settings in S3}\label{tab3}	
		\begin{tabular}{cc}
			\hline
			Parameters & Specification \\ \hline
			Scan Time & $\mathrm{T}=10$ s \\
			Meas. Noise Cov. & $\mathbf C_s^v = \mathrm{diag}(1,1)$ \\
			Multi. Noise Cov. & $\mathbf C^h =  \frac{1}{3} \mathbf I_2$ \\
			Cov. w.r.t Kine. & $\mathbf C^x_w = \mathrm{diag}(50,50,1,1)$ \\
			Cov. w.r.t Extent & $\mathbf C^p_w = \mathrm{diag}(0.05,0.01,0.001)$ \\
			Prior Cov. w.r.t Kine. & $\mathbf {C}_{1,s}^{x[0]} = \mathrm{diag} (50,50, 1,1)$ \\
			Prior Cov. w.r.t Extent & $\mathbf {C}_{1,s}^{p[0]} = \mathrm{diag} (0.36, 5, 5)$ \\
			\hline
		\end{tabular}
	\end{minipage}
\end{figure}  

Fig. \ref{ellip_gwd} shows the GWDs on the examined filters. For two types of centralized filters, CEOT gives a better result since the uncertainty of semi-axes lengths is modeled by low noise intensity, and the orientation is described by high noise intensity. The CEOT-RM only uses a single constant $\tau$ to describe the degrees of freedom about the extent, resulting in a larger GWD value.

In S2, DEOT is no longer valid as it works properly only if each node is full-view at any scan time. Therefore, the GWD in DEOT is always large than that of CM and CI. For the two proposed filters, CM outperforms CI and is nearly the same as in CEOT. In fact, CM needs infinite iterations so that the consensus weights $\pi_L^{s,j} \to \frac{1}{\lvert \mathcal G \rvert}$. Then, CM converges to CEOT with $\omega_{k,s} = \lvert \mathcal G \rvert$. However, a trade-off must be made on the choice of iteration $L$ to leverage computation burden and performance. Here, we set $L=6$ and choose a more suitable $\omega_{k,s}$ as given in \cite{b25} to eliminate the inconsistency between CEOT and CM. The means ensure the superiority of CM.  

It is worth noting that CM is more sensitive to the number of measurements. The conclusion is confirmed in Fig. \ref{ellip_gwd} by comparing two cases (i.e., the number of measurements follows a Poisson distribution with means $\lambda=5$ and $\lambda=10$). Moreover, we observe from Fig. \ref{ellip_avegwd} that the GWD in CM has a decreased tendency with more iterations (see the results on nodes $10$ and $16$). In contrast, CI has a faster convergence rate even by a single iteration and is less sensitive to more measurements. Hence, according to the theoretical analysis and simulation results, a reasonable anticipation is that as the number of iterations increases, CM will exhibit lower GWD error.

As shown in Figs. \ref{ellip_acee1}-\ref{ellip_acee}, CI and CM using the corresponding consensus schemes maintain a small bias in the kinematics and extent. Here, DEOT is a compromise of the two filters in the extent, but the ACEE value is much higher in the kinematics especially during the turning process. This is because DEOT using a diffusion technique only computes a combination of the neighboring estimates with equal weight to remain consensus. With increased iterations, better consensus (lower ACEE value) is achieved as shown in Figs. \ref{ellip_aveacee1}-\ref{ellip_aveacee}. It is also important to note that CM has a decreased trend at the first few iterations while CI has a slight fluctuation.

Fig. \ref{ellip_com} gives the average computational cost per tracking process under different iterations. We observe from Fig. \ref{ellip_com} that the required computation resource is limited and nearly identical in CM and CI, which meets the real-time requirement of a tracking scene.

\subsubsection{Rectangular with Nearly Constant Velocity Model Scenario}
\label{sub:7.2.2}
In this scenario (S3), the object is a rectangle with lengths of the semi-axes $10$m and $5$m. The object moves with nearly constant speed $v=50 \mathrm{km/h}$ following the trajectory as shown in Fig. \ref{tra}. The parameters used in the examined filters are listed in Table \ref{tab3}. The comparison results are the OSPA distance, ACEE and computation cost over $M=50$ Monte Carlo runs.

\begin{figure}
	\begin{minipage}[b]{0.45\linewidth}
		\centering
		\includegraphics[width=3.2in,height=2.3in]{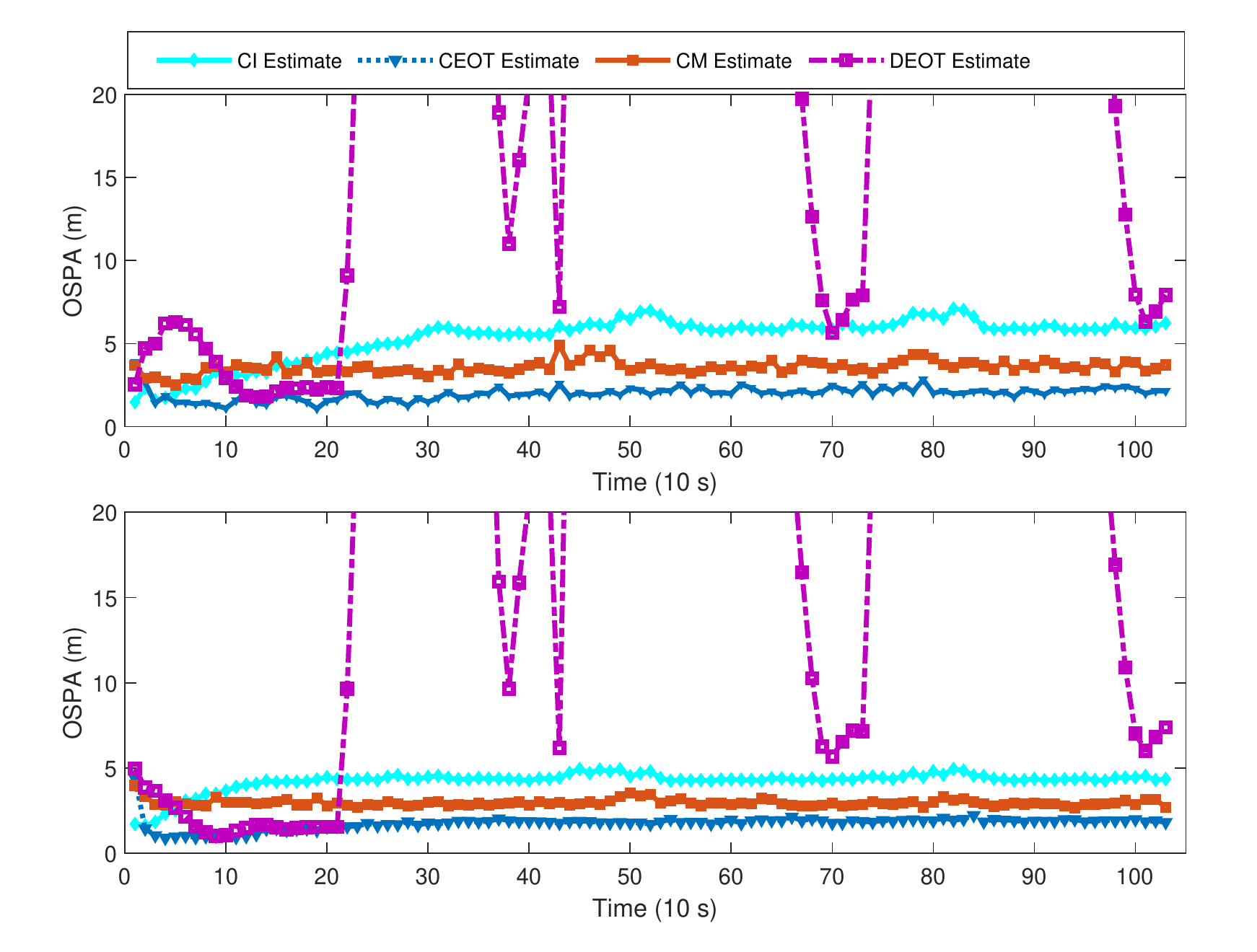}
		\caption{OSPAs using CEOT, DEOT, CM, and CI with iteration $L=6$. The first row shows the OSPAs under $\lambda=5$. The second row shows the GWDs under $\lambda=10$.}
		\label{rec_ospa}
	\end{minipage}
	\begin{minipage}[b]{0.55\linewidth}
		\centering
		\includegraphics[width=3.2in,height=2.3in]{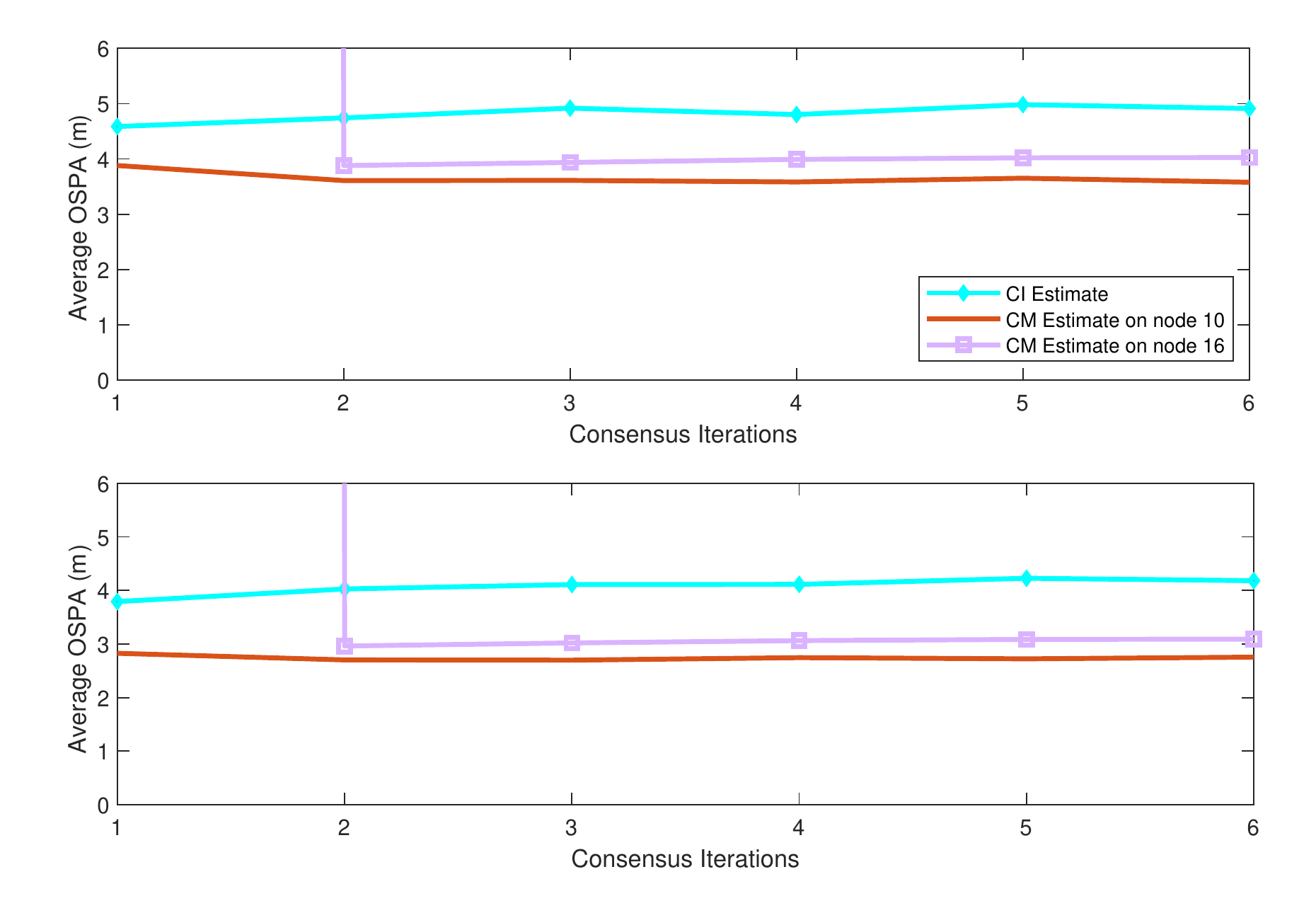}
		\caption{Average OSPAs with different iterations. The first row shows the Average OSPAs under $\lambda=5$. The second row shows the Average OSPAs under $\lambda=10$.}
		\label{rec_aveospa}
	\end{minipage}
\end{figure}

The OSPAs  are shown in Fig. \ref{rec_ospa} under different number of measurements ($\lambda=5$ and $\lambda=10$). From Fig. \ref{rec_ospa}, the OSPA error in DEOT is much higher than that of CI and CM. It is seen that CI does not follow CEOT as closely as in CM, especially for a low number of measurements. Combined with the results in Figs. \ref{rec_ospa}-\ref{rec_aveospa}, we observe that CM has increased performance with more iterations or measurements, but CI is slightly affected by the two aspects. 

\begin{figure}
	\begin{minipage}[thpb]{0.45\linewidth}
		\centering
		\includegraphics[width=3.2in,height=2.3in]{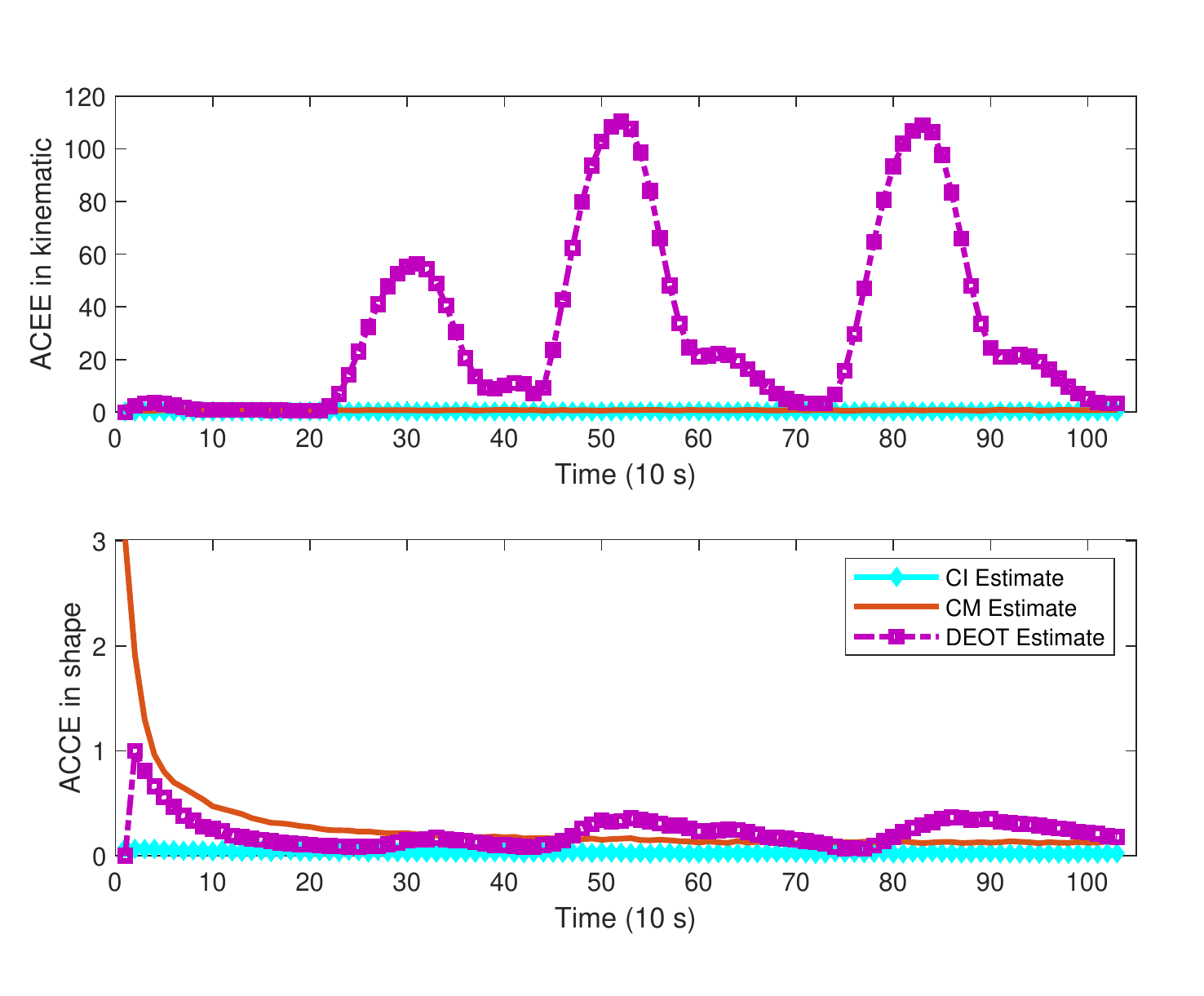}
		\caption{ACEEs with iteration $L=6$ ($\lambda=5$).}
		\label{rec_acee1}
	\end{minipage}
	\begin{minipage}[thpb]{0.55\linewidth}
		\centering
		\includegraphics[width=3.2in,height=2.3in]{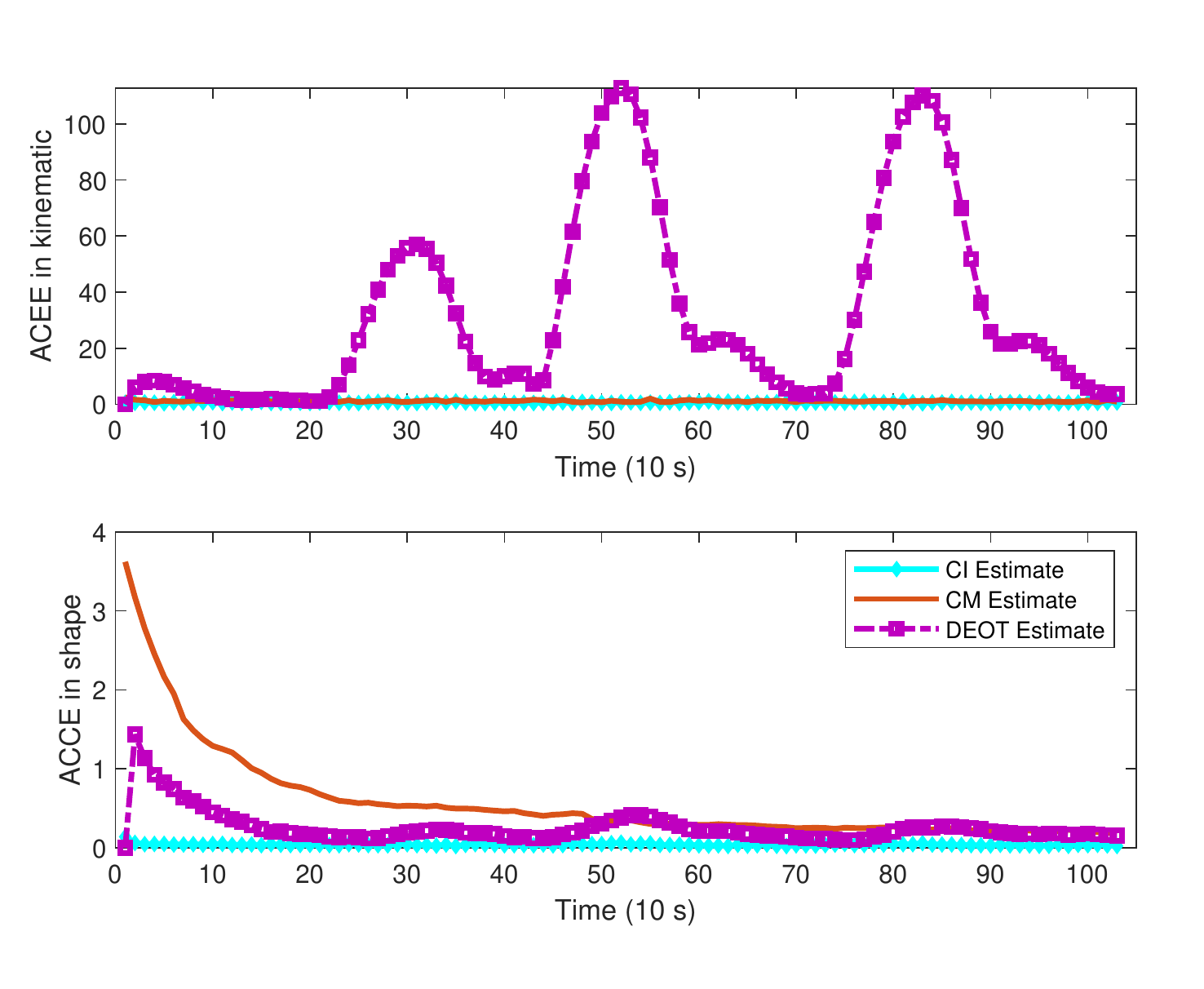}
		\caption{ACEEs with iteration $L=6$ ($\lambda=10$).}
		\label{rec_acee}
	\end{minipage}
\end{figure}

\begin{figure}
	\begin{minipage}[thpb]{0.45\linewidth}
		\centering
		\includegraphics[width=3.2in,height=2.3in]{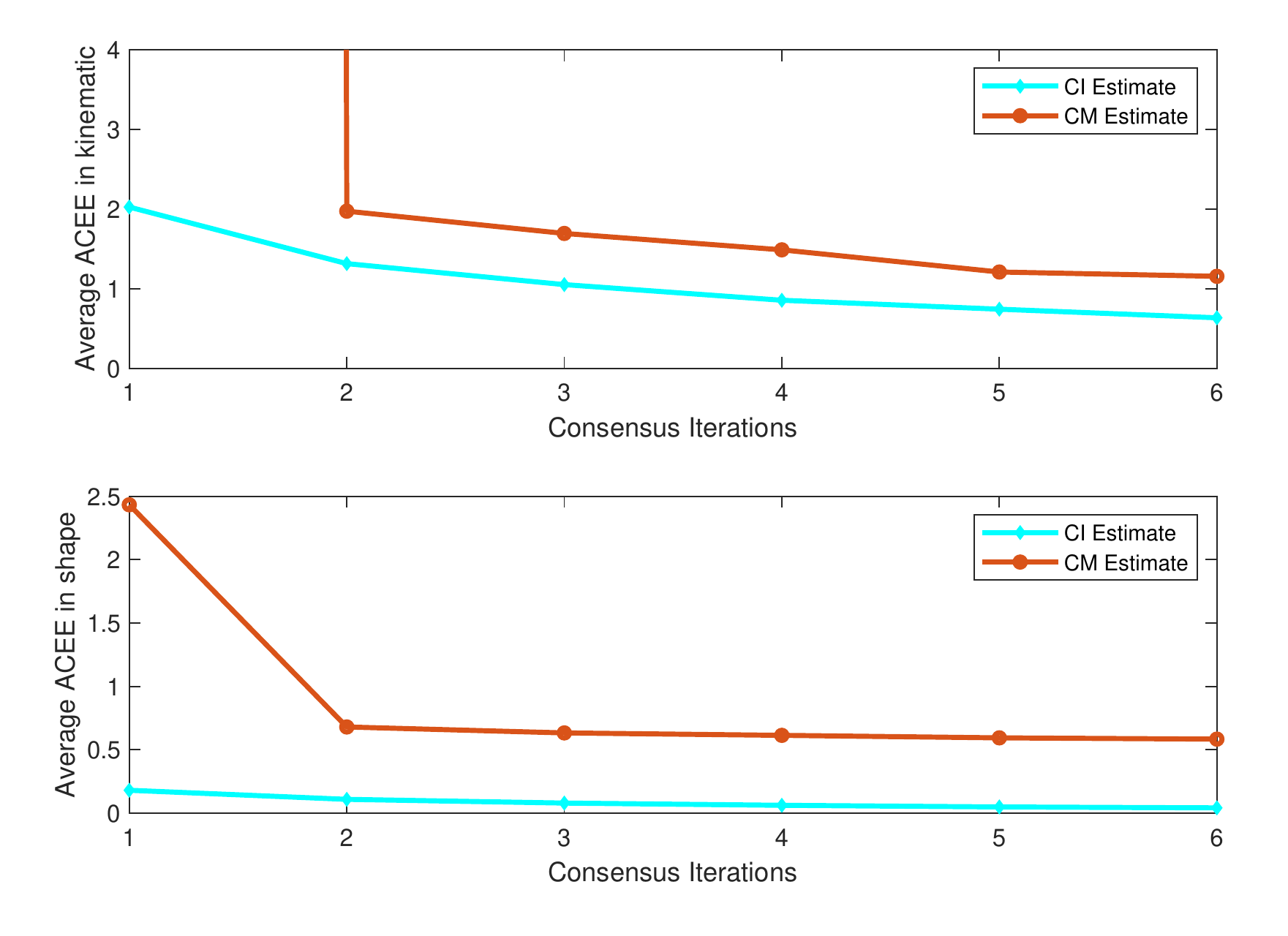}
		\caption{Average ACEEs with different iterations under $\lambda=5$.}
		\label{rec_aveacee}
	\end{minipage}
	\begin{minipage}[thpb]{0.55\linewidth}
		\centering
		\includegraphics[width=3.3in,height=2.3in]{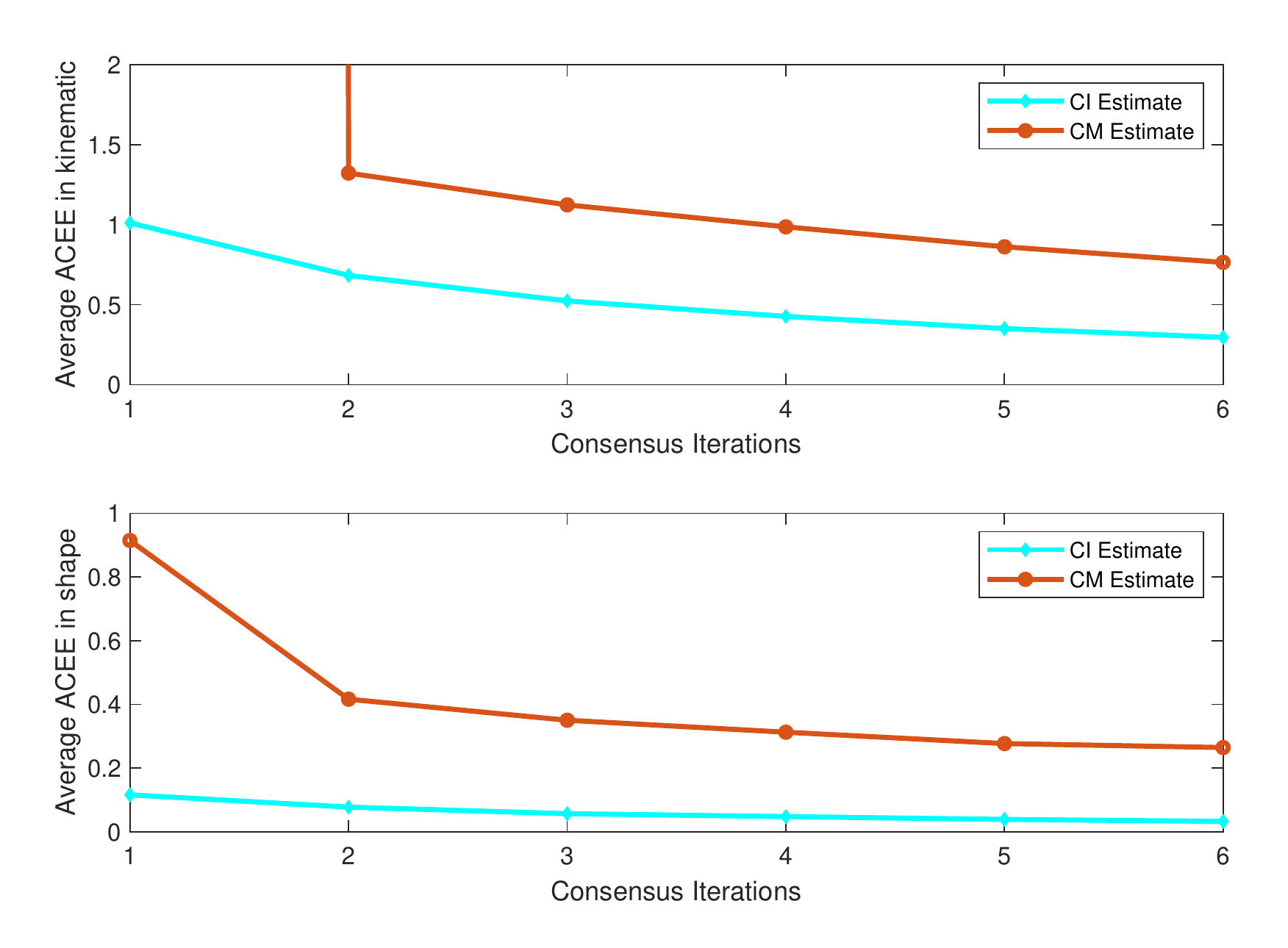}
		\caption{Average ACEEs with different iterations under $\lambda=10$.}
		\label{rec_aveacee1}
	\end{minipage}
\end{figure}

As expected, CI and CM give a satisfactory consensus result in the kinematics and extent as shown in Figs. \ref{rec_acee1}-\ref{rec_acee}. However, DEOT using the diffusion technique lacks the ability to reduce the estimation bias in a practical network. The average ACEE decreases with increased iterations as shown in Figs. \ref{rec_aveacee}-\ref{rec_aveacee1}, and the trend becomes more obvious in CM during a few iterations.

\begin{figure}
	\centering
	\includegraphics[scale=.6]{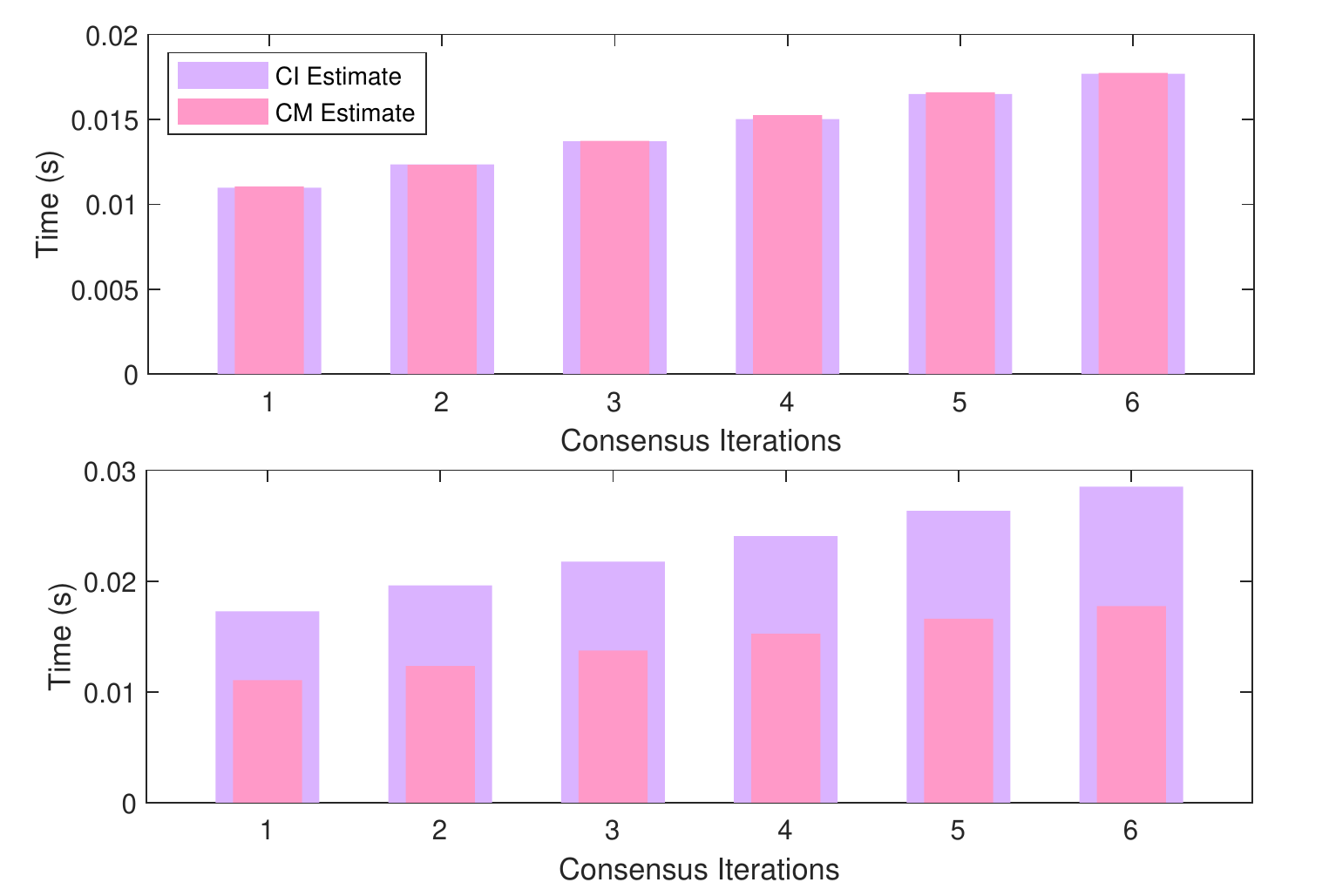}
	\caption{Average computational time (CT) under different iterations. The first row shows the average CT under $\lambda = 5$, and the second shows the average CT under $\lambda = 10$.}
	\label{rec_com}
\end{figure}

Fig. \ref{rec_com} shows the average computational cost per tracking process with the number of iterations set to $1$ to $6$. For any given $l$, CM almost consumes identical computation resources as its counterpart CI, which expands their practicability in a real scenario. 

\subsection{Evaluation on the consistency}
In addition to tracking error and estimation bias, an efficient filter is also required to evaluate the consistency. A common measure of consistency is the NEES,
\begin{equation*}
	\textrm{NEES}_x=\frac{1}{M} \sum\limits_{m=1}^M ({\hat {\bm x}}_{k,s}^{m} - \bm x_k)^{\mathsf{T}} (\mathbf C_{k|k}^{m})^{-1}({\hat {\bm x}}_{k,s}^{m} - \bm x_k)
\end{equation*}
where the superscript $m$ denotes the estimate for trial $m$. The NESS of a consistent filter should be close to the dimension of estimated state $n$ (here, $n=4$ for kinematic, and $n=3$ for extent), and be chi-square distributed with $nM$ degrees of freedom. 

The objective of this experiment is to evaluate the consistency of CI and CM. To this end, we choose a part of time steps in the case described in section \ref{sub:7.2.1} to generate true kinematics and extent. All of the parameters are the same as in Table \ref{tab2}.  

\begin{figure}
	\begin{minipage}[thpb]{0.45\linewidth}
		\centering
		\includegraphics[width=3.2in,height=2.3in]{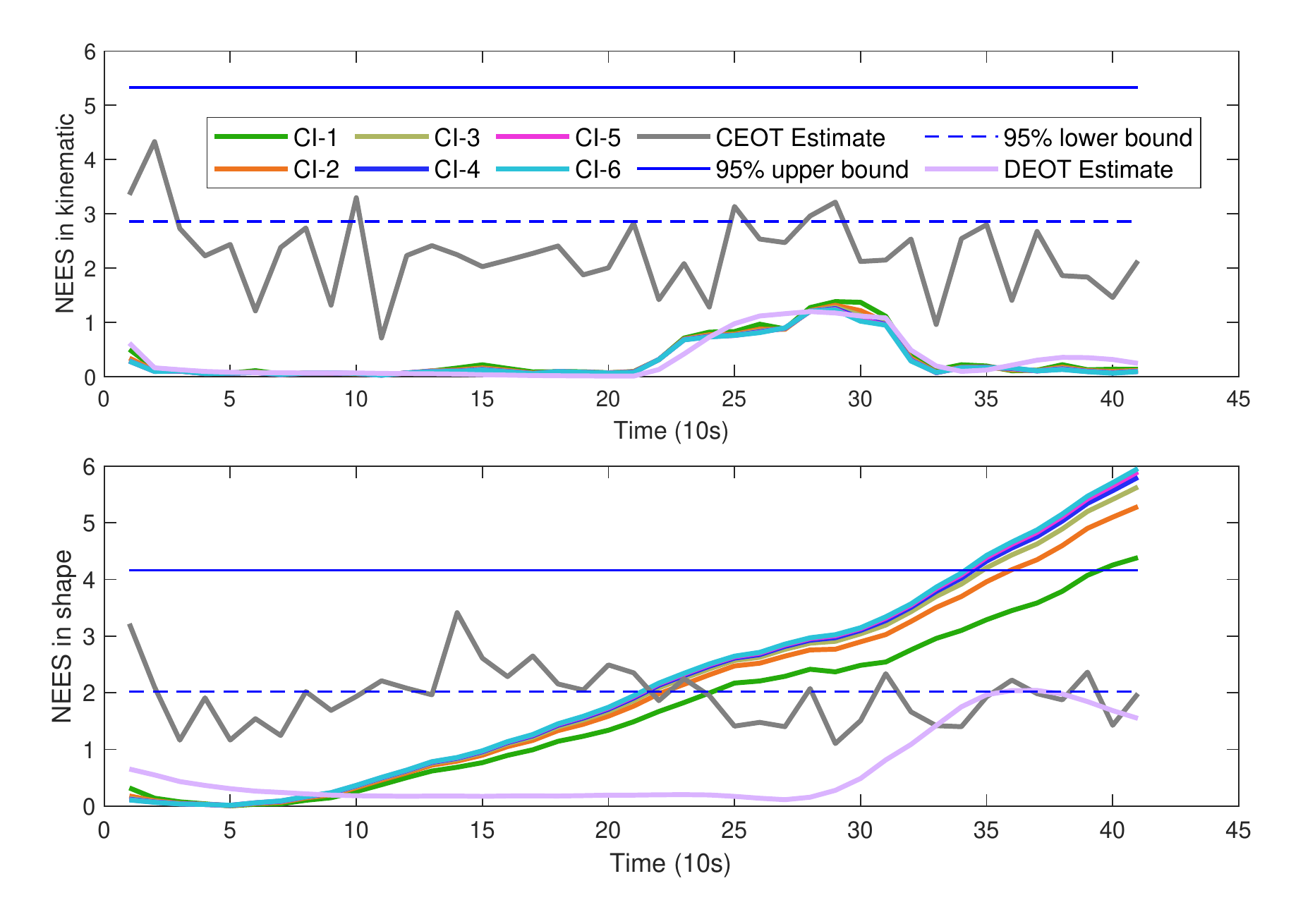}
		\caption{NESSs in CEOT, CI, and DEOT under $\lambda=10$.}
		\label{ness_CI}
	\end{minipage}
	\begin{minipage}[thpb]{0.55\linewidth}
		\centering
		\includegraphics[width=3.2in,height=2.3in]{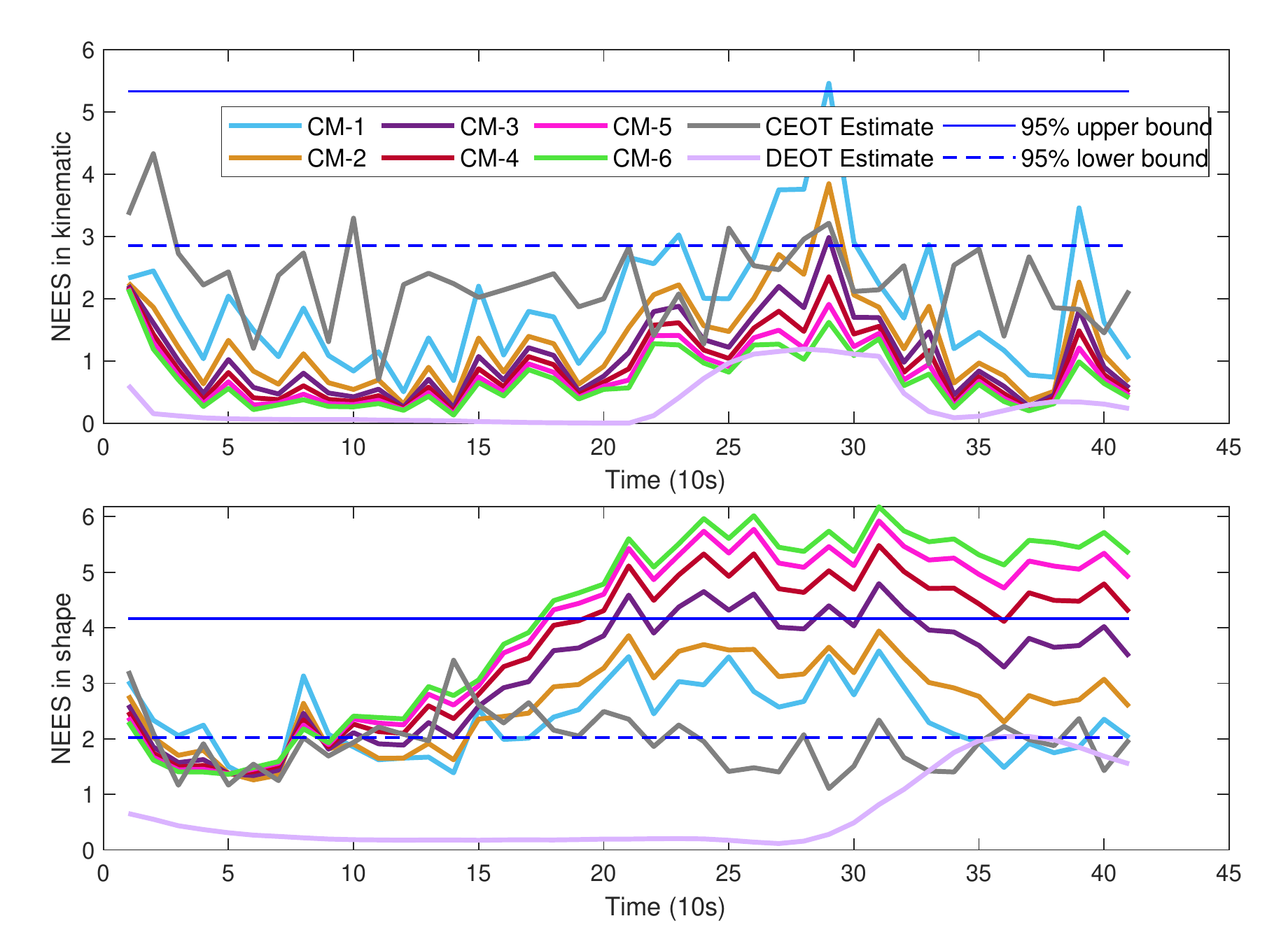}
		\caption{NESSs in CEOT, CM, and DEOT under $\lambda=10$.}
		\label{ness_CM}
	\end{minipage}
\end{figure}

The NEESs for CM and CI with different $L$ are shown in Figs. \ref{ness_CI} and \ref{ness_CM}, respectively. Here, $L$ is varied from $1$ to $6$ at increments of $1$. Meanwhile, the NEESs for CEOT and DEOT are also given as a reference. For both NEESs about the kinematics and extent, we see from Fig. \ref{ness_CI} that the most values of NEESs in CEOT approach to the lower bound of $95\%$ confidence interval. But in DEOT, both NEESs lie lower than the confidence interval, which shows poor consistency of the local estimates. From Fig. \ref{ness_CI}, the NEESs in CI for the kinematics follow a similar trend in DEOT even for different $L$, while the NESSs for the extent vary towards the upper bound and have approximately the same magnitude for $L>2$.  

Results in Fig.\ref{ness_CM} exhibit that CM has better consistency overall than that of CI for both the kinematics and extent. And its superiority is more apparent to the extent when $L$ is relatively small since the NEES lies within the confidence interval. 

\section{Conclusion}
\label{sec:conclusion}
In this work, dual linear state-space models, without losing the cross-correlation between states, are designed as a basis to meet the information filter style. Following this, a centralized and two types of distributed extended object tracking information filters, namely CI and CM, are developed. Moreover, it is proven that the estimation errors of the proposed filter are bounded in the mean square. Numerical results exhibit the superiority of the proposed filters in contrast to current distributed works. 

This work mainly focuses on a scenario that multiple sensors observe a 2-D object from different field-of-view. However, the extent of extended objects in the real environment must be three-dimensional, and thus the objective of fusion is how to use multiple sensors to get the 3-D kinematics and extent. This research deserves further exploration but out of the scope of the work. 

Potential future works are devoted to investigating extensions of the two pair linear measurement models, e.g., state estimation over time-varied networks \cite{b34}, heterogeneous networks \cite{b38}, and state estimation under multiple constraints \cite{b39,b40}. 
\section*{Acknowledgment}
This work is supported by the National Natural Science Foundation of China (Grants no. 61873205, 62273283 and 61771399).

\section*{Appendix A}
 \label{appdix1}
\begin{proofof}[Proof of Proposition \ref{exteup}.]
	Notice that \eqref{eq11} does not give a direct mapping between $\bm p_k$ and $\mathbf Y_{k,s}^{i}$. Thus, we substitute \eqref{eq3} and \eqref{eq9} into \eqref{eq11} to get the expression, 
	\begin{subequations} \label{eqA1}
		\begin{align}
			\mathbf Y_{k,s}^{i} \approx &{} \mathbf F \left( \left( \mathbf H \bm x_k - \mathbf H \hat{\bm x}_k^{[i-1]} + \hat{\mathbf{S}}_k^{[i-1]} \bm{h}_{k,s}^{i}+\left[\begin{array}{c}
				\left(\bm{h}_{k,s}^{i}\right)^{\mathsf{T}} \hat{\mathbf{J}}_{1,k}^{[i-1]} \\
				\left(\bm{h}_{k,s}^{i}\right)^{\mathsf{T}} \hat{\mathbf{J}}_{2,k}^{[i-1]}
			\end{array}\right]\left(\bm{p}_k-\hat{\bm{p}}_k^{[i-1]}\right) + \bm  v_{k,s}^{i}\right) \otimes (\star) \right) \label{eqA1a} \\
			:= &{} [Y_1^2\: Y_2^2 \: Y_1Y_2]^{\mathsf{T}}. \label{eqA1b}
		\end{align}
	\end{subequations}
	With the above results, the goal becomes to generate a linear measurement model $\mathbf Y_{k,s}^{i} \approx \mathbf H^p_k \bm p_k + \bm v_{k,s}^{p[i]}$ with respect to $\bm p_k$ from \eqref{eqA1a}, where $\mathbf H^p_k$ denotes the measurement matrix and $\bm v_{k,s}^{p[i]}$ is the measurement noise. Meanwhile, its first and second moments are desired to match the expectation and covariance of \eqref{eq11} as much as possible. We first define $\mathbf H \bm x_k - \mathbf H \hat{\bm x}_k^{[i-1]} := [\tilde{x}_{1}\;\tilde{x}_{2}]^{\mathsf{T}}$, $\mathbf H \mathbf C_k^{x[i-1]} \mathbf H^{\mathsf{T}} := \left[\begin{smallmatrix}
		c^x_{11} & c^x_{12} \\ c^x_{21} & c^x_{22}
	\end{smallmatrix}\right] $, $\mathbf C_{k,s}^{y[i]} :=\left[\begin{smallmatrix}
		c^y_{11} & c^y_{12} \\ c^y_{21} & c^y_{22}
	\end{smallmatrix}\right]$, $\hat{\mathbf{S}}_k^{[i-1]} := [\hat{\mathbf{S}}_{1}^{\mathsf{T}}\;\hat{\mathbf{S}}_{2}^{\mathsf{T}}]^{\mathsf{T}}$, $\bm v_{k,s}^{i}:=[v_{1}\;v_{2}]^{\mathsf{T}}$, $\mathrm{Cov} (\bm v_{k,s}^{i}) := \mathrm{diag}\left[\begin{smallmatrix}
		\sigma_1^2 & 0 \\ 0 & \sigma_2^2
	\end{smallmatrix} \right] $ and omit the superscript $[\cdot]$, time index $k$, and sensor node index $s$.
	Then, we further expand \eqref{eqA1b} as below 
	\begin{subequations} \label{eqA2}
		\begin{equation} \label{eqA2a}
			\begin{split}
				Y_1^2  = \;& \tilde{x}_{1}^2 + (\hat{\mathbf{S}}_{1}\bm{h})^{\mathsf{T}} (\star) + \left( \bm{h}^{\mathsf{T}} \hat{\mathbf{J}}_{1} (\bm{p}-\hat{\bm{p}})\right) ^{\mathsf{T}} (\star) + v_{1}^2 + 2 \tilde{x}_{1} \hat{\mathbf{S}}_{1} \bm h + 2 \tilde{x}_{1} \bm h^{\mathsf{T}} \hat{\mathbf{J}}_{1} (\bm{p}-\hat{\bm{p}}) + 2 \tilde{x}_{1} v_{1} + 2 \hat{\mathbf{S}}_{1} \bm h \bm{h}^{\mathsf{T}} \hat{\mathbf{J}}_{1} (\bm{p}-\hat{\bm{p}}) \\
				& + 2 \hat{\mathbf{S}}_{1} \bm h v_{1}  + 2 \bm h^{\mathsf{T}} \hat{\mathbf{J}}_{1} (\bm{p}-\hat{\bm{p}}) v_{1} - 2 \hat{\mathbf{S}}_{1} \mathbf C^h \hat{\mathbf{J}}_{1} \bm{p} +  2 \hat{\mathbf{S}}_{1} \mathbf C^h \hat{\mathbf{J}}_{1} \bm{p}
			\end{split} 
		\end{equation} 
		\begin{equation} \label{eqA2b}
			\begin{split}
				Y_2^2  = \;& \tilde{x}_{2}^2 + (\hat{\mathbf{S}}_{2}\bm{h})^{\mathsf{T}} (\star) + \left( \bm{h}^{\mathsf{T}} \hat{\mathbf{J}}_{2} (\bm{p}-\hat{\bm{p}})\right) ^{\mathsf{T}} (\star) + v_{2}^2 + 2 \tilde{x}_{2} \hat{\mathbf{S}}_{2} \bm h + 2 \tilde{x}_{2} \bm h^{\mathsf{T}} \hat{\mathbf{J}}_{2} (\bm{p}-\hat{\bm{p}}) + 2 \tilde{x}_{2} v_{2} + 2 \hat{\mathbf{S}}_{2} \bm h \bm{h}^{\mathsf{T}} \hat{\mathbf{J}}_{2} (\bm{p}-\hat{\bm{p}}) \\
				& + 2 \hat{\mathbf{S}}_{2} \bm h v_{2}  + 2 \bm h^{\mathsf{T}} \hat{\mathbf{J}}_{2} (\bm{p}-\hat{\bm{p}}) v_{2} - 2 \hat{\mathbf{S}}_{2} \mathbf C^h \hat{\mathbf{J}}_{2} \bm{p} +  2 \hat{\mathbf{S}}_{2} \mathbf C^h \hat{\mathbf{J}}_{2} \bm{p}
			\end{split} 
		\end{equation}
		\begin{equation} \label{eqA2c}
			\begin{split}
				Y_1Y_2 = \;& \tilde{x}_{1} \tilde{x}_{2} + (\hat{\mathbf{S}}_{1}\bm{h})^{\mathsf{T}} (\hat{\mathbf{S}}_{2}\bm{h}) + \left( \bm{h}^{\mathsf{T}} \hat{\mathbf{J}}_{1} (\bm{p}-\hat{\bm{p}})\right) ^{\mathsf{T}} \left( \bm{h}^{\mathsf{T}} \hat{\mathbf{J}}_{2} (\bm{p}-\hat{\bm{p}})\right) + v_1v_2 + 2 \tilde{x}_{1} \hat{\mathbf{S}}_{2} \bm h + 2 \tilde{x}_{1} \bm h^{\mathsf{T}} \hat{\mathbf{J}}_{2} (\bm{p}-\hat{\bm{p}}) + 2 \tilde{x}_{1} v_2 \\
				& + 2 \hat{\mathbf{S}}_{1} \bm h \bm{h}^{\mathsf{T}} \hat{\mathbf{J}}_{2} (\bm{p}-\hat{\bm{p}}) + 2 \hat{\mathbf{S}}_{1} \bm h v_2 + 2 \bm h^{\mathsf{T}} \hat{\mathbf{J}}_{1} (\bm{p}-\hat{\bm{p}}) v_{2} + \hat{\mathbf{S}}_{2} \mathbf C^h \hat{\mathbf{J}}_{1} \bm{p} +
				\hat{\mathbf{S}}_{1} \mathbf C^h \hat{\mathbf{J}}_{2} \bm{p} - \hat{\mathbf{S}}_{2} \mathbf C^h \hat{\mathbf{J}}_{1} \bm{p} -
				\hat{\mathbf{S}}_{1} \mathbf C^h \hat{\mathbf{J}}_{2} \bm{p}.
			\end{split} 
		\end{equation}
	\end{subequations}
	The expectation of \eqref{eqA2} is given as follows:
	\begin{subequations} \label{eqA3}
		\begin{equation} \label{eqA3a}
			\mathbb{E} (Y_1^2) = c_{11}^x + \hat{\mathbf{S}}_{1} \mathbf C^h \hat{\mathbf{S}}_{1}^{\mathsf{T}} + \mathrm{tr} (\mathbf C^p \hat{\mathbf{J}}_{1}^{\mathsf{T}} \mathbf C^h \hat{\mathbf{J}}_{1}) + \sigma_1^2 - 2 \hat{\mathbf{S}}_{1} \mathbf C^h \hat{\mathbf{J}}_{1} \bm{p} +  2 \hat{\mathbf{S}}_{1} \mathbf C^h \hat{\mathbf{J}}_{1} \bm{p} 
		\end{equation}
		\begin{equation} \label{eqA3b}
			\mathbb{E} (Y_2^2) = c_{22}^x + \hat{\mathbf{S}}_{2} \mathbf C^h \hat{\mathbf{S}}_{2}^{\mathsf{T}} + \mathrm{tr} (\mathbf C^p \hat{\mathbf{J}}_{2}^{\mathsf{T}} \mathbf C^h \hat{\mathbf{J}}_{2}) + \sigma_2^2 - 2 \hat{\mathbf{S}}_{2} \mathbf C^h \hat{\mathbf{J}}_{2} \bm{p} +  2 \hat{\mathbf{S}}_{2} \mathbf C^h \hat{\mathbf{J}}_{2} \bm{p} 
		\end{equation}
		\begin{equation} \label{eqA3c}
			\mathbb{E} (Y_1 Y_2) = c_{12}^x + \hat{\mathbf{S}}_{1} \mathbf C^h \hat{\mathbf{S}}_{2}^{\mathsf{T}} + \mathrm{tr} (\mathbf C^p \hat{\mathbf{J}}_{1}^{\mathsf{T}} \mathbf C^h \hat{\mathbf{J}}_{2}) + \hat{\mathbf{S}}_{2} \mathbf C^h \hat{\mathbf{J}}_{1} \bm{p} +
			\hat{\mathbf{S}}_{1} \mathbf C^h \hat{\mathbf{J}}_{2} \bm{p} - \hat{\mathbf{S}}_{2} \mathbf C^h \hat{\mathbf{J}}_{1} \bm{p} -
			\hat{\mathbf{S}}_{1} \mathbf C^h \hat{\mathbf{J}}_{2} \bm{p}. 
		\end{equation}
	\end{subequations}
	The equation \eqref{eqA2a} implies $\mathbb{E} (Y_1) = \mathbb{E} (Y_2) =0$, $\mathrm{Cov} (Y_1) = c_{11}^y$, $\mathrm{Cov} (Y_1Y_2) = c_{12}^y$ and $\mathrm{Cov} (Y_2) = c_{22}^y$. According to Wick's theorem \cite{b41}, we get 
	\begin{equation} \label{eqA4}
		\left \{ \begin{lgathered}
			\mathbb{E} \{(Y_1^2)^2\} = 3 (c_{11}^y)^2, \; \mathbb{E} \{(Y_2^2)^2\} = 3 (c_{22}^y)^2, \; \mathbb{E} \{(Y_1^3Y_2)\} = 3 (c_{11}^y c_{12}^y) \\ \mathbb{E} \{(Y_1 Y_2^3)\} = 3 (c_{22}^y c_{12}^y), \; \mathbb{E} \{(Y_1^2 Y_2^2)\} =  c_{11}^y c_{22}^y + 2 (c_{12}^y)^2
		\end{lgathered} \right.. 
	\end{equation} 
	By rearranging \eqref{eqA2b}, \eqref{eqA3}, and \eqref{eqA4} yields 
	\begin{equation} \label{eqA5}
		\left[\begin{array}{c} Y_1^2  \\ Y_2^2 \\ Y_1 Y_2 \end{array}\right] = \hat{\mathbf M} \bm p + \bm v^p 
	\end{equation}
	with expectation $\mathbf F \mathrm{vect} \left(\mathbf C^{y}\right)$ and covariance $\mathbf F \left( \mathbf C^{y}\otimes \mathbf C^{y} \right) (\mathbf F + \tilde{\mathbf F})^{\mathsf{T}}$. Then, we get the linear model as shown in \eqref{eq13}, and find the measurement matrix $\mathbf H^p = \hat{\mathbf M}$. These two moments of \eqref{eqA5} are equal to the expectation and covariance of \eqref{eq11}, respectively, which means that \eqref{eqA2} achieves the moment matching. In fact, for the last few terms of \eqref{eq30}, other terms can be chosen to make the equation hold. Choosing those terms in $\hat{\mathbf M}$ ensures \eqref{eq11} and \eqref{eqA1a} giving an identical result on computing the cross-covariance between $\mathbf Y$ and $\bm p$, while using other terms cannot obtain the result. The proof is complete.
	
\end{proofof}
\setcounter{equation}{0}
\renewcommand\theequation{B.\arabic{equation}}
\begin{proofof}[Proof of Theorem \ref{them1}.]
	Define the prediction and estimation errors at node $s \in \mathcal{G}$ as $\bm e_{k+1|k,s} = \bm x_{k+1} - \hat{\bm x}_{k+1|k,s}$ and $\bm e_{k,s} = \bm x_{k} - \hat{\bm x}_{k|k,s}$, respectively. Denote the collective errors $\bm e_{k+1|k} = \mathrm{col}(\bm e_{k+1|k,s}, s \in \mathcal{G})$ and $\bm e_{k} = \mathrm{col}(\bm e_{k|k,s}, s \in \mathcal{G})$.
	
	Let $\bm \tau:=[\tau_1,\cdots,\tau_{\lvert \mathcal G \rvert}]$ be the Perron-Frobenius left eigenvector of the matrix $\mathbf \Pi^L:=(\pi_L^{s,j})_{\lvert \mathcal G \rvert \times \lvert \mathcal G \rvert}$. The components of the eigenvector are strictly positive and has $\sum_{j \in \mathcal G} \tau_j \pi_L^{s,j} = \tau_s$. 
	
	Consider the following stochastic function about $e_{k+1|k}$
	\begin{equation} \label{eqB1}
		V(\bm e_{k+1|k}) = \sum_{s \in \mathcal{G}} \tau_{s}(\bm e_{k+1|k,s})^{\mathsf{T}} \mathbf{\Omega}_{k+1|k,s}^{x} \bm e_{k+1|k,s},
	\end{equation}
	where $\bm e_{k+1|k,s}$ follows that
	\begin{equation} \label{eqB2}
		\begin{split}
			\bm e_{k+1|k,s} = \;& \bm x_{k+1} - \hat{\bm x}_{k+1|k,s} = \mathbf F_k^x (\bm x_{k} - \hat{\bm x}_{k|k,s}) + \bm w_k^x \\
			= & \mathbf F_k^x  (\mathbf{\Omega}_{k|k,s}^{x})^{-1} \mathbf{\Omega}_{k|k-1,s}^{x} (\bm x_k - \hat{\bm x}_{k|k-1,s}) - \sum_{j \in \mathcal G^{s}} \pi_L^{s,j} \omega_{k,s} \mathbf F_k^x (\mathbf{\Omega}_{k|k,s}^{x})^{-1} (\bm{\beta}_{k,j} \mathbf H)^{\mathsf{T}} \mathbf V_{k,j}^{x}  \bm{v}_{k,j}^x + \bm w_k^x.
		\end{split}
	\end{equation}
	According to the Metropolis weights rule used in CM filter,
	\begin{equation} \label{eqB3}
		\begin{split}
			\sum_{j\in\mathcal G} \pi_L^{s,j} \mathbf{\Omega}_{k|k-1,j}^{x} & (\bm x_k -  \hat{\bm x}_{k|k-1,j}) \\ =\;& (1 - \sum_{j \in \mathcal G, j \neq s} \pi_L^{s,j})\: \mathbf{\Omega}_{k|k-1,s}^{x} (\bm x_k - \hat{\bm x}_{k|k-1,s}) + \sum_{j \in \mathcal G, j \neq s} \pi_L^{s,j} \mathbf{\Omega}_{k|k-1,j}^{x} (\bm x_k - \hat{\bm x}_{k|k-1,j})\\
			=\; & \mathbf{\Omega}_{k|k-1,s}^{x} (\bm x_k - \hat{\bm x}_{k|k-1,s}) + \sum_{j \in \mathcal G, j \neq s} \pi_L^{s,j} \left[ \mathbf{\Omega}_{k|k-1,j}^{x} (\bm x_k - \hat{\bm x}_{k|k-1,j}) - \mathbf{\Omega}_{k|k-1,s}^{x} (\bm x_k - \hat{\bm x}_{k|k-1,s})\right] .
		\end{split}
	\end{equation}
	Since the predicted information matrix and information vector are performed in local nodes, the former updated estimates govern the difference in predicted estimates. A fact is that the term $\mathbf{\Omega}_{k|k-1,s}^{x} (\bm x_k - \hat{\bm x}_{k|k-1,s})$ in \eqref{eqB2} approaches to $\sum_{j\in\mathcal G^s} \pi_L^{s,j} \mathbf{\Omega}_{k|k-1,j}^{x} (\bm x_k - \hat{\bm x}_{k|k-1,j})$ under the dual influence of iteration operation and compensation factor $\omega_{k,s}^{[i]}$. Using the fact, \eqref{eqB2} is rewritten as  
	\begin{equation} \label{eqB4}
		\bm e_{k+1|k,s} \approx \mathbf F_k^x  (\mathbf{\Omega}_{k|k,s}^{x})^{-1} \left[ \sum_{j\in\mathcal G} \pi_L^{s,j} \mathbf{\Omega}_{k|k-1,j}^{x} (\bm x_k - \hat{\bm x}_{k|k-1,j}) - \sum_{j\in\mathcal G} \pi_L^{s,j} \omega_{k,s} (\bm{\beta}_{k,j} \mathbf H)^{\mathsf{T}} \mathbf V_{k,j}^{x}  \bm{v}_{k,j}^x \right]  + \bm w_k^x.
	\end{equation}
	Next, inserting \eqref{eqB4} to \eqref{eqB1} and taking the conditional expectation yields 
	\begin{equation} \label{eqB5}
		\mathbb{E} \{ V(\bm e_{k+1|k})| \bm e_{k|k-1} \} = \Phi_{k+1}^x + \Phi_{k+1}^v + \Phi_{k+1}^w,
	\end{equation}
	where
	\begin{equation} \label{eqB6}
		\left \{ \begin{lgathered}
			\Phi_{k+1}^x = \mathbb{E} \left[ \sum_{s \in \mathcal{G}} \tau_s (\star)^{\mathsf{T}}\mathbf{\Omega}_{k+1|k,s}^{x} \left( \sum_{j \in \mathcal{G}} \pi_L^{s,j}\mathbf F_k^x  (\mathbf{\Omega}_{k|k,s}^{x})^{-1} \mathbf{\Omega}_{k|k-1,j}^{x} \bm e_{k|k-1,s}\right) | \bm e_{k|k-1,s}  \right]   \\
			\Phi_{k+1}^v = - \mathbb{E} \left[ \tau_s (\star)^{\mathsf{T}}\mathbf{\Omega}_{k+1|k,s}^{x} ( \sum_{j \in \mathcal{G}} \pi_L^{s,j} \omega_{k,s} \mathbf F_k^x  (\mathbf{\Omega}_{k|k,s}^{x})^{-1} (\bm{\beta}_{k,j} \mathbf H)^{\mathsf{T}} \mathbf V_{k,j}^{x}  \bm{v}_{k,j} ) | \bm e_{k|k-1,s} \right] \\
			\Phi_{k+1}^w = \mathbb{E} \left[ \tau_s (\star)^{\mathsf{T}}\mathbf{\Omega}_{k+1|k,s}^{x} \bm w_k^x| \bm e_{k|k-1,s} \right] 
		\end{lgathered} \right.. 
	\end{equation}
	Next, the proof of boundedness about the three terms in \eqref{eqB5} is similar to that of Theorem 1 in \cite{neuro,hybird18}. The detailed proof can refer to  \cite{neuro,hybird18}. 
\end{proofof}

\bibliography{mybibfile}

\end{document}